\documentclass[%
 reprint,
 amsmath,amssymb,
 aps,
pra,
]{revtex4-2}

\usepackage{siunitx}
\usepackage{verbatim}
\usepackage{graphicx}
\usepackage{dcolumn}
\usepackage{bm}
\usepackage{float}
\usepackage{hyperref}
\usepackage{mathtools}
\usepackage{svg}

\newcommand{\nc}{n_\mathrm{c}}
\newcommand{\nin}{n_\mathrm{i}}
\newcommand{\nf}{n_\mathrm{f}}
\newcommand{\np}{n_\mathrm{p}}
\newcommand{\nm}{n_\mathrm{m}}

\newcommand{\om}{\omega_{\rm m}}
\newcommand{\oo}{\omega_{\rm o}}
\newcommand{\ola}{\omega_{\rm l}}

\newcommand{\gz}{\gamma_\mathrm{0}}

\newcommand{\gm}{\gamma_\mathrm{m}}
\newcommand{\gp}{\gamma_\mathrm{p}}
\newcommand{\gaom}{\gamma_\mathrm{om}}

\newcommand{\ke}{\kappa_\mathrm{e}}
\newcommand{\ki}{\kappa_\mathrm{i}}

\newcommand{\gom}{g_\mathrm{om}}

\DeclareUnicodeCharacter{2009}{\,} 

\begin{document}

\preprint{APS/123-QED}

\title{Optomechanical crystal in light-resilient quantum ground state}

\author{Johan Kolvik$^1$}
\author{Paul Burger$^1$}
\author{David Hambraeus$^1$}
\author{Trond H. Haug$^1$}
\author{Joey Frey$^1$}
\author{Mads B. Kristensen$^1$}
\author{Raphaël Van Laer$^{1,}$}%
\email{raphael.van.laer@chalmers.se}
\affiliation{%
 $^1$Department of Microtechnology and Nanoscience (MC2), Chalmers University of Technology. 41298 Göteborg, Sweden}%

\date{\today}

\begin{abstract}
Interaction between light and high-frequency sound is a key area in integrated photonics, quantum and nonlinear optics, and quantum science. However, typical suspended optomechanical structures suffer from poor thermal anchoring, making them susceptible to thermal noise arising from optical absorption. Here, we demonstrate a chip-scale, \textit{release-free} silicon optomechanical crystal cavity operating cryogenically with improved resilience to laser light. Relative to a suspended design, we observe \qty{18}{\decibel} suppression of the thermo-optic effect, and the device also sustains near-unity phonon occupation at \qty{35}{\decibel} higher intracavity optical energy in continuous-wave operation. 
Non-exponential decay dynamics from an as yet unidentified thermal process limit transfer of this performance to pulsed operation. Resolving this channel would open a clear path to near-term quantum protocols, such as microwave-to-optical quantum transduction. More broadly, the release-free architecture offers a novel platform for studying thermal noise dynamics with results relevant across silicon optomechanical platforms.
\end{abstract}

\maketitle


\section*{Introduction} \label{sec:introduction}
Nanomechanical systems have emerged as a promising platform for both classical and quantum technologies \cite{safavi-naeini_controlling_2019}.
The mechanical system can act as a universal bus, efficiently bridging otherwise weakly coupled degrees of freedom such as microwave and optical photons \cite{zou_microwave-optical_2021}. 
The silicon optomechanical crystal (OMC) is a compelling gigahertz-frequency nanomechanical system that over the past decade has enabled several key advances, including active ground-state cooling \cite{chan_laser_2011,mayor_high_2025,qiu_laser_2020}, microwave-optical entanglement \cite{jiang_optically_2023,meesala_non-classical_2024,meesala_quantum_2024}, remote mechanical entanglement \cite{riedinger_remote_2018}, generation of non-classical states of light \cite{riedinger_non-classical_2016,hong_hanbury_2017,chen_low-noise_2026}, and mechanical states with second-scale lifetimes \cite{maccabe_nano-acoustic_2020}. 
These devices are attractive in the quantum regime, thanks to their strong interaction between light and GHz sound and, in the absence of light, passive initialization in the mechanical ground-state at millikelvin temperatures. 

However, further progress is limited by elevated thermal noise arising from optical absorption, a challenge shared across MHz- \cite{huang_room-temperature_2024} and GHz-frequency \cite{meenehan_silicon_2014,meenehan_pulsed_2015} suspended optomechanics. While the original and most common OMC designs use suspended nanobeams to achieve low mechanical dissipation, this also results in thermally isolated structures vulnerable to build-up of thermal noise. Although two-dimensional suspended devices designed to mitigate heating demonstrate promise in both pulsed and continuous-wave operation \cite{ren_two-dimensional_2020, sonar_high-efficiency_2025, mayor_high_2025}, their noise performance remains a challenge in applications combining optical and microwave photons \cite{xie_two-dimensional_2026}. Another promising approach -- although more challenging to scale -- involves macroscopic bulk mechanical modes more resistant to laser heating \cite{doeleman_brillouin_2023}.
These limitations highlight the need to revisit the basic structural design principles governing the interaction between light and high-frequency sound to enable faster and higher-fidelity quantum optomechanical operations. 

In an effort to further improve thermal anchoring in nano-scale optomechanical devices, we recently demonstrated \textit{release-free} optomechanical crystal cavities that maintain strong mechanical confinement and efficient light-sound interaction even without suspension \cite{kolvik_clamped_2023,burger_design_2025}. 
This new class of OMCs leverages mechanical modes that are intrinsically protected from radiating into surface and bulk acoustic waves through total internal reflection. 
At room temperature, these release-free devices showed encouraging optomechanical zero-point coupling rates of $\gom/(2\pi) = \qty{0.5}{\mega\hertz}$ and optical (mechanical) quality factors of $2.4\cdot 10^5$ (850) \cite{kolvik_clamped_2023}.

\begin{figure*}
    \centering
    \includegraphics{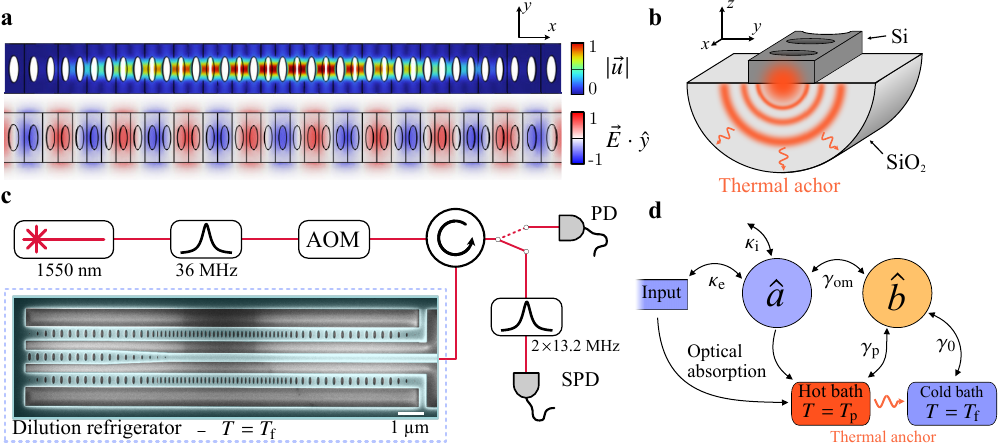}
    \caption{\textbf{Release-free optomechanical crystal for improved thermal anchoring}. \textbf{a} Top-down view of simulated mechanical and optical mode profiles showing normalized displacement $\vec{u}$ and electrical $\vec{E}$ fields respectively. The optomechanical crystal (OMC) consists of a defect region with 31 unit cells between two adiabatically tapered mirror regions \cite{kolvik_clamped_2023}. \textbf{b} The bottom side of the silicon nanobeam is fully attached to the underlying substrate, opening up a channel through which heat can decay. The mechanical mode of interest is protected from leakage by total internal reflection. \textbf{c} Simplified phonon-counting setup and a scanning electron micrograph of two release-free OMCs adjacent to an optical bus waveguide. We place the devices in a dilution refrigerator at temperature $T_\mathrm{f}$ and study them using a near-infrared laser with a wavelength around \qty{1550}{\nano\meter}. Using acousto-optic modulation (AOM), we send both continuous-wave and pulsed signals to the devices via a circulator. Finally, we measure the reflected light with a photodiode (PD) or through single-photon detection (SPD). The SPD arm singles out optomechanical sidebands by removing pump light with two filter cavities with \qty{13.2}{\mega\hertz} bandwidth. \textbf{d} Phenomenological model describing the mechanical environment under optical pumping. The optical mode $\hat{a}$ is populated through an input waveguide with rate $\ke$ and decays to the environment through internal losses $\ki$. Pumping the optical mode gives rise to the optomechanical interaction that allows phonon-photon exchange at rate $\gaom = 4\gom^2\nc/\kappa$ where $\kappa =\ki + \ke$ and $\nc$ is the number of intracavity pump photons. The mechanical mode $\hat{b}$ is coupled to the hot (cold) thermal bath at rates $\gp$ ($\gz$) where the cold bath is thermalized to $T_\mathrm{f}$ and the hot bath to an elevated temperature $T_\mathrm{p}$ due to optical absorption in $\hat{a}$ and the input waveguide. Enhanced thermal anchoring opens a direct channel for hot bath phonon decay into the cold bath, reducing the thermal load on the mechanical mode.} 
    \label{fig:release-free concept}
\end{figure*}

In this work, we investigate the cryogenic performance of the release-free OMCs, focusing on optical power handling and associated thermal noise. 
By benchmarking against conventional suspended nanobeam OMCs, we show that the release-free design improves resistance to heating resulting from optical absorption. We begin to show that the thermo-optic effect in our release-free devices activates at around \qty{18}{\decibel} higher optical intracavity energy compared to the suspended counterpart.
We further characterize thermal noise through single-phonon-counting, finding that the release-free device supports optical pump powers up to \qty{35}{\decibel} higher than the suspended version when operated at thermal noise occupancy $\nm < 1$.  
The noise's time dynamics follow non-trivial behavior resulting from an as yet unidentified source. This deviation from exponential decay currently limits the device's performance in pulsed operation. Despite this, the steady-state light-resilience allows for an average of $<0.5$ noise phonons for \qty{100}{\nano\second} pulse trains with \qty{100}{\kilo\hertz} repetition rate at \qty{5}{\%} scattering probability. Our results position release-free optomechanics as a possible path toward low-noise silicon optomechanics in addition to high power classical acousto-optics.

\vspace*{-1\baselineskip} 
\section*{Results}
\subsection*{Release-free optomechanical crystals}  \label{sec:release-free OMCs}
The release-free optomechanical crystal confines GHz sound through the principle of total internal reflection \cite{kolvik_clamped_2023}. 
The mechanical mode of interest resides near the X-point of the defect-unit-cell's band structure, featuring a pinching-type motion (Fig.~\ref{fig:release-free concept}a). 
Due to geometrical softening \cite{lagasse_higherorder_1973, kolvik_clamped_2023}, the choice of substrate need not be restricted to materials with higher bulk sound velocity than for the silicon device layer. 
This enables release-free OMC designs compatible with a wide range of platforms including silicon-on-insulator (SOI) \cite{kolvik_clamped_2023} and silicon-on-sapphire \cite{burger_design_2025}, such that fabrication can be tailored to the needs of the application. 
The present device is fabricated on SOI with a \qty{220}{\nano\meter} device layer \cite{kolvik_clamped_2023} on \qty{2}{\micro\meter} thermal silicon dioxide. 
Having an entire side of the device in full contact with the chip surface opens up a large area through which thermal noise can dissipate (Fig.~\ref{fig:release-free concept}b). 
An increased rate of thermal phonon dissipation should allow the OMC to operate at higher applied optical power at a given thermal noise phonon occupancy $\nm$.

To analyze the thermal noise in our devices, we consider a phenomenological model for the mechanical mode's thermal environment (Fig.~\ref{fig:release-free concept}d) \cite{meenehan_silicon_2014}. 
The mechanical mode $\hat{b}$ couples to an optical cavity $\hat{a}$ through the parametrically-enhanced optomechanical scattering rate $\gaom$ (Fig.~\ref{fig:release-free concept}). In addition to the two coupled oscillators, the model also introduces two thermal baths that couple incoherently to the mechanical mode. The cold bath is the collection of modes that thermalize to the temperature of the surrounding environment $T_\mathrm{f}$ and interacts with the mechanical mode with the intrinsic decay rate $\gz$. The hot bath on the other hand, is conjectured to be a thermal bath populated by high-frequency phonons thermalized to an elevated temperature $T_\mathrm{p}$ \cite{maccabe_nano-acoustic_2020}. This elevated temperature likely results from optical absorption in electronic defect states present in the etched silicon surfaces \cite{borselli_measuring_2006}.  Through nonlinear interactions, the high-frequency thermal phonons would then downconvert and mix with the protected GHz mechanical mode at a rate $\gp$ \cite{sonar_high-efficiency_2025}. Optical absorption occurs in all etched silicon surfaces implying that hot bath stimulus can come from both the OMC active region and indirectly via the optical bus waveguide supplying the OMC with light (Fig.~\ref{fig:release-free concept}c,d) \cite{ren_two-dimensional_2020,sonar_high-efficiency_2025}.

A central aspect of the release-free geometry is the material interface between the thin-film silicon and the underlying substrate. 
The release-free approach relies on the protected mechanical mode being confined on the thin-film side of the interface, whereas thermal noise phonons are not. 
However, due to the mismatch of material properties at the interface, even the noise phonons may reflect partially \cite{chen_interfacial_2022,zeng_phonon_2000} resulting in interfacial thermal resistance (ITR). 
Although theoretical models exist, there appears to be little experimental work on ITR in thin-film structures at cryogenic temperatures. This makes it an open question whether release-free devices can indeed deliver improved thermal handling, or whether their performance is limited by ITR.

In this study, we compare the thermalization properties of a release-free OMC \cite{kolvik_clamped_2023} to that of a conventional suspended nanobeam OMC  \cite{chan_optimized_2012}. The devices are fabricated identically apart from that the suspended OMCs undergo a final suspension step before measurement (App.~\ref{app:fabrication}). The release-free and suspended devices are fixed to the mixing chamber of a dilution refrigerator to allow characterization under varying ambient temperature (App.~\ref{app:optical_setup}). Before delving into the thermal properties, we characterize the optomechanical performance of the devices (App.~\ref{app:dev prop}). We begin by studying the optical properties by laser spectroscopy. For the release-free OMC, we measure total optical linewidth $\kappa/(2\pi)=\qty{1.63}{\giga\hertz}$, and external coupling $\ke/(2\pi)=\qty{680}{\mega\hertz}$. Next, we measure the optomechanical coupling and mechanical frequency at a dilution fridge temperature of 10 mK. Our cryogenic electromagnetically-induced-transparency (EIT) measurements (App.~\ref{app:eit}) \cite{safavi-naeini_electromagnetically_2011} yield a mechanical frequency $\om/(2\pi) = \qty{5.358}{\giga\hertz}$ and optomechanical coupling $\gom/(2\pi)=\qty{470}{\kilo\hertz}$ for the release-free device. 

\begin{figure}
    \centering
    \includegraphics{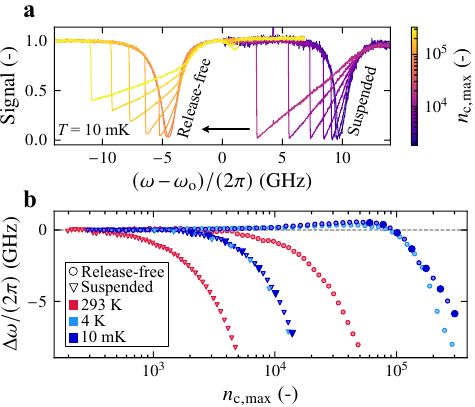}
    \caption{\textbf{Suppressed thermo-optic nonlinearity in a release-free optomechanical crystal.} \textbf{a} Normalized optical reflection recorded at 10 mK where release-free and suspended data is horizontally offset for visual clarity. We record the data while sweeping the laser frequency in a sawtooth pattern with direction indicated by the black arrow. For each optical power we calculate the maximum intracavity photon number $n_\mathrm{c,max}$ achieved at resonant pumping. \textbf{b} Optical frequency shift $\Delta\omega$ as function of temperature and intracavity photons $\nc$. At each power, we extract the optical resonance frequency shift. Bold data points correspond to the curves presented in \textbf{a}.}
    \label{fig:thermo-optic}
\end{figure}

\subsection*{Thermo-optic robustness} 
In order to investigate the OMCs' thermal properties, we first study the thermo-optic nonlinearity as a simple and approximate indicator of the thermal anchoring. When an OMC heats up due to optical absorption, the silicon's temperature-dependent refractive index causes the optical resonance to redshift \cite{barclay_nonlinear_2005}. We study this phenomenon by sweeping a tunable laser (Fig.~\ref{fig:thermo-optic}a) over the OMC optical mode and recording the resonance frequency shift $\Delta\omega$ as a function of the per-sweep maximally attained photon occupation $n_\mathrm{c,max}$. Taking into account the strong temperature dependence of the OMC material's thermal properties, we perform the experiment at three different dilution fridge temperatures (Fig.~\ref{fig:thermo-optic}b). 

\begin{figure*}
    \centering
    \includegraphics{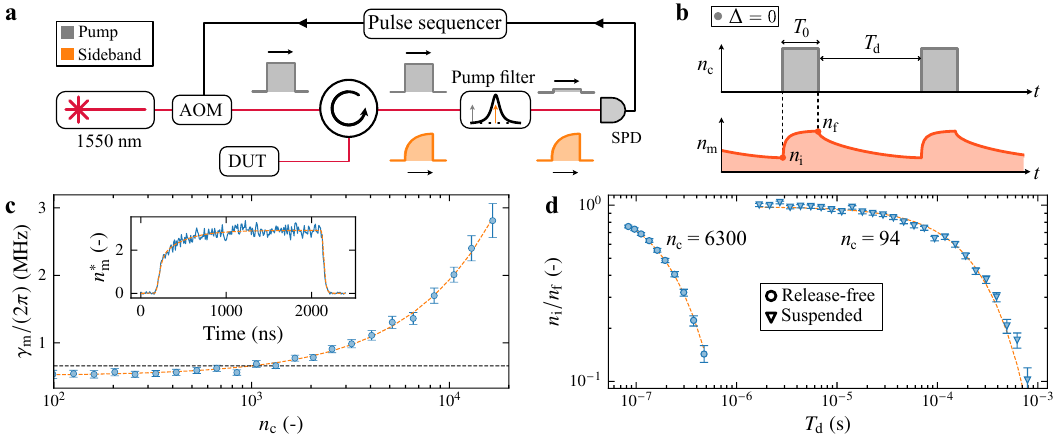}
    \caption{
        \textbf{Thermal noise dynamics of resonantly-driven release-free and suspended OMCs.} \textbf{a} Simplified schematic of the experimental setup used for pulsed phonon-counting. We illuminate the device under test (DUT) with high extinction ratio optical pulses generated through acousto-optic modulation (AOM). Next, we use a set of Fabry--Pérot filters to single out the anti-Stokes sideband at frequency $+\om$ relative to the pump. We handle timing between optical pulsing and single-photon detection (SPD) using a pulse sequencer. \textbf{b} The pulse sequence: We drive and read out with resonant pulses of duration $T_\text{0}$ and delay $T_\text{d}$. Using Eq.~\eqref{eq:thermal_noise_dynamics}, we extract initial $(\nin)$ and final $(\nf)$ phonon occupation during the optical illumination.  \textbf{c} Total mechanical linewidth for the release-free device as function of $\nc$. We study the dynamics of the scattered optomechanical sideband when illuminating the device with $T_0 = \qty{2}{\micro\second}$. For each optical power of the experiment, we fit the collected SPD data (see inset) with Eq.~\eqref{eq:thermal_noise_dynamics} while accounting for filter dispersion to extract the total linewidth with least squares standard deviation. The estimated linewidth agrees with the dark ($\nc = 0$) ringdown result in \textbf{d} (black dashed line). Lastly, we fit the linewidth data with a power law (orange dashed line) $\gm(\nc) = \gz + \gp(\nc),$ where $\gp(\nc) = b(\nc)^a$. \textbf{d} Normalized initial phonon occupation versus delay time $T_\text{d}$ with 95\% credibility intervals. We measure exponential decay for both devices, revealing light-off decay rates $\gz/(2\pi) = \qty{660}{\kilo\hertz}$ $(\qty{528}{\hertz})$ for the release-free (suspended) device.
    }
    \label{fig:pulse_exp}
\end{figure*}

At room-temperature, a strong thermo-optic redshift dominates the nonlinear behavior of both suspended and release-free devices. However, we see that the release-free device withstands an order of magnitude higher intracavity optical photons at fixed redshift. Now, we repeat the experiment at cryogenic temperatures. The thermo-optic nonlinearity diminishes drastically at lower temperatures \cite{komma_thermo-optic_2012} which is indicated by the onset of redshifting at higher optical cavity occupancy for both devices (Fig.~\ref{fig:thermo-optic}b). At both 4 K and 10 mK, we measure a greatly increased power stability for the release-free device which now show equal redshift to the suspended device for at least 18 dB increased intracavity photons. We observe additional nonlinearities in the release-free device that activate at $n_\mathrm{c,max}>10^4$. First of these nonlinearities is a blueshift measured in the range $n_\mathrm{c,max}\in[10^4,10^5]$ which we attribute to free carrier dispersion \cite{barclay_nonlinear_2005}. In addition, the high optical occupancies achieved in the release-free device also cause nonlinear losses resulting in diminishing optical extinction ratios (Fig.~\ref{fig:thermo-optic}a). An absorptive loss channel such as two-photon absorption would cause additional heat in the OMC, further driving the thermo-optic effect \cite{barclay_nonlinear_2005}. Finally, we observe an optomechanical sideband at $+\om$ relative to the optical frequency at high optical cavity occupancy $n_\mathrm{c,max}$, which we attribute to mechanical self-oscillations. 

The nonlinear behavior of the devices is practically unchanged between 4 K and 10 mK. We attribute this to the vanishing thermal capacity of the devices at low temperatures and the low dilution fridge cooling power at 10 mK. At the high pumping powers required for appreciable nonlinear response, the devices equilibrate at the same final temperature regardless of their initial temperature. Due to the vanishing thermo-optic coefficient at low temperatures \cite{komma_thermo-optic_2012} we would also see no difference in the thermo-optic response which strengthens this interpretation. Lastly, the increased optical power tolerance of the release-free device indicates that the ITR between the thin-film silicon and the underlying thermal SiO$_2$ does not severely limit the thermal capacity of the device even at cryogenic temperatures.

\subsection*{Hot and cold bath coupling}  \label{sec:bath dynamics}
To study the noise properties of our devices, we return to the phenomenological model in Fig.~\ref{fig:release-free concept}d. The model suggests that thermal noise in the mechanical mode is governed by its interaction with the hot and cold thermal baths. In this section, we study these interactions at a dilution fridge temperature of 10 mK through pulsed single-phonon-counting with a resonant (detuning $\Delta = 0$) optical pump (Methods). We use acousto-optic modulation to generate high-extinction-ratio optical pulses, and use pump filters to selectively measure the anti-Stokes sideband on our SPD (Fig.~\ref{fig:pulse_exp}a). For resonant pumping, optomechanical back-action is suppressed which allows for isolated study of the hot and cold bath dynamics $(\gm=\gz+\gp)$ \cite{aspelmeyer_cavity_2014}. In addition, we reach higher thermal phonon occupancy at resonant pumping due to higher achievable intracavity optical powers. We use this to study noise dynamics at intermediate phonon occupation levels $(\nm \approx 10)$. To calibrate the setup, we use sideband asymmetry measured through SPD (App.~\ref{app:calibration}). 

When exciting the release-free device with \qty{2}{\micro\second} long resonant square pulses, we measure sideband waveforms with significantly longer rise-times than the input pulses (Fig.~\ref{fig:pulse_exp}b). The increased rise time stems from a thermally driven ring-up of the mechanical mode caused by optical absorption in the pump pulse. Assuming that the mechanical mode is coupled to a hot bath at an elevated temperature during optical excitation, we describe the thermal occupation dynamics with
\begin{equation}
    \label{eq:thermal_noise_dynamics}
    \nm(t) = \nin + (\nf - \nin)(1 - e^{-\gm t}),\quad t\in[0,T_\text{0}].
\end{equation}
Here, $\nin$ is the initial phonon occupation at the start of the pulse and $\nf$ is the occupation reached for pulse durations $T_\mathrm{0}$ far exceeding the total mechanical decay rate $\gm$. In addition to the dynamics of the mechanical mode, the strong dispersion of the pump filters also leads to distortion in the pulse shape measured by the SPD. We therefore take care to distinguish between the actual time-dependent phonon occupation $\nm(t)$ and the measured $\nm^*(t)$ in our analysis (App.~\ref{app:filter_dynamics},\ref{app:bayesian_fitting}).

Using the noise dynamics model in Eq.~\eqref{eq:thermal_noise_dynamics}, we now study the release-free OMC's heating rate at varying optical powers. We fit the measured phonon occupation to Eq.~\eqref{eq:thermal_noise_dynamics} while considering the impact of the filter transfer function to extract $\gm(\nc)$ (Fig.~\ref{fig:pulse_exp}c). We see an increase in $\gm$ for higher intracavity photons $\nc$, and we fit the trend to a power law $\gm(\nc) = \gz + \gp(\nc),$ where $\gp(\nc) = b(\nc)^a$. From the fit we extract parameters $\gz/(2\pi) = \qty{510}{\kilo\hertz}$, $b/(2\pi) = \qty{170}{\hertz}$ and $a = 0.98$. We also find that this $\gp(\nc)$ dependence appears in EIT measurements (App.~\ref{app:eit}). The exponent parameter $a$ is significantly higher than the previously measured 0.29 for two-dimensional devices at high $\nc$ \cite{ren_two-dimensional_2020, sonar_high-efficiency_2025}. This higher measured exponent is compatible with a three-dimensional ansatz for the hot bath density of states, which predicts $a = 1$ for $T_\text{p}\gg\hbar\om/k_\text{B}$ \cite{maccabe_nano-acoustic_2020}.

In addition, we measure the mechanical decay rate in absence of light by monitoring the ringdown of the thermal mechanical noise for both release-free and suspended devices. We run the test with equal pulse parameters to the previous experiment while setting the power to $\nc = 6300$ $(\nc = 94)$ for the release-free (suspended) device. We choose the powers of the experiment to achieve $\nf\approx10$ for both devices 
. Returning to Eq.~\eqref{eq:thermal_noise_dynamics}, we extract the initial and final phonon occupation during pulsing and study their dependence on the pulse delay time $T_\text{d}$ (Fig.~\ref{fig:pulse_exp}d). Both devices show exponential decay of the relative initial phonon occupation $\nin/\nf$ with time constants $\gz/(2\pi) = \qty{660}{\kilo\hertz}$ $(\qty{528}{\hertz})$  for the release-free (suspended) device (Fig.~\ref{fig:pulse_exp}d). For the release-free device, this value agrees with the low-power result in Fig.~\ref{fig:pulse_exp}c and with earlier EIT measurements (App.~\ref{app:eit}).

The large difference in cold-bath coupling between the devices primarily arises from their design. In previous work, we showed that the release-free design exhibits a radiation-limited quality factor of approximately $10^5$, substantially lower than that achievable in properly shielded suspended nanobeams \cite{kolvik_clamped_2023,maccabe_nano-acoustic_2020}. The measured quality factor of the release-free device is further reduced by approximately \SI{10}{\decibel} relative to this limit, likely due to fabrication imperfections that induce spurious phonon scattering. A stronger cold-bath coupling $\gz$ has a nuanced impact on device performance. While it reduces the maximum achievable single-photon cooperativity, it simultaneously enhances the coupling of the mechanical mode to the cryostat cold bath, thereby reducing susceptibility to the absorption-induced hot bath (Fig.~\ref{fig:release-free concept}d). Consistent with this trade-off, suspended devices designed for low-noise operation are often intentionally engineered to exhibit reduced mechanical quality factors comparable to that of the present release-free device \cite{patel_engineering_2017,riedinger_remote_2018}. The implications of stronger cold-bath coupling for relevant applications are discussed in App.~\ref{app:transducer_efficiency}.

\begin{figure}
    \centering
    \includegraphics{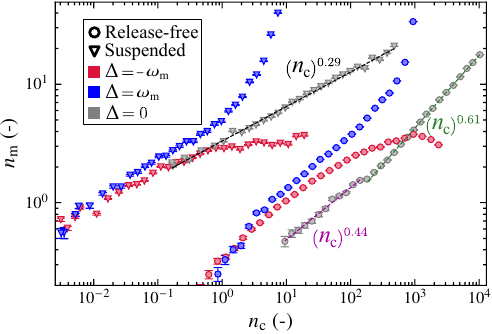}
    \caption{\textbf{Single-phonon operation at high continuous-wave optical power at 10 mK.} Average mechanical occupation under continuous-wave optical illumination in release-free and suspended OMCs for red, blue and resonant pumping. Data is given with one standard deviation assuming Poissonian statistics for single-photon-counting. To the resonant data, we fit power laws of the form $\nm(\nc) = \alpha \cdot(\nc)^\beta$, where the different exponents are highlighted in the figure.  
    }
   
    \label{fig:CW_n_vs_nc}
\end{figure}

\subsection*{Continuous-wave noise thermometry} \label{sec:cw}
With further knowledge of the hot and cold bath couplings $(\gz,\gp)$, we move on to study the thermal occupation of our devices under continuous-wave operation. We extract the average mechanical occupation $\nm(\nc)$ as function of intracavity photons $\nc$ for red, blue and resonant pumping (Fig.~\ref{fig:CW_n_vs_nc}). As expected, thermal noise generally increases with higher $\nc$ for both devices. In addition, off-resonant pumping $(\Delta=\pm\om)$ yields phonon cooling and amplification due to optomechanical back-action when $\gaom$ approaches $\gz+\gp$. For resonant pumping, these effects are suppressed leaving thermal driving as the sole contributor to the measured noise occupancy. The mechanical occupancy $\nm$ in Fig.~\ref{fig:CW_n_vs_nc} represents the total noise in the mechanical mode and differs from the occupation of the hot phonon bath $\np$. In the case of the suspended OMC, we expect a dominant coupling to the hot bath due to the low $\gz$ (Fig.~\ref{fig:pulse_exp}d) which yields $\nm \approx \np$ \cite{maccabe_nano-acoustic_2020}. However, this assumption does not apply to the release-free device which appears to be predominantly coupled to the cold bath in the range $\nc < 3500$ (Fig.~\ref{fig:pulse_exp}c).

\begin{figure*}
    \centering
    \includegraphics{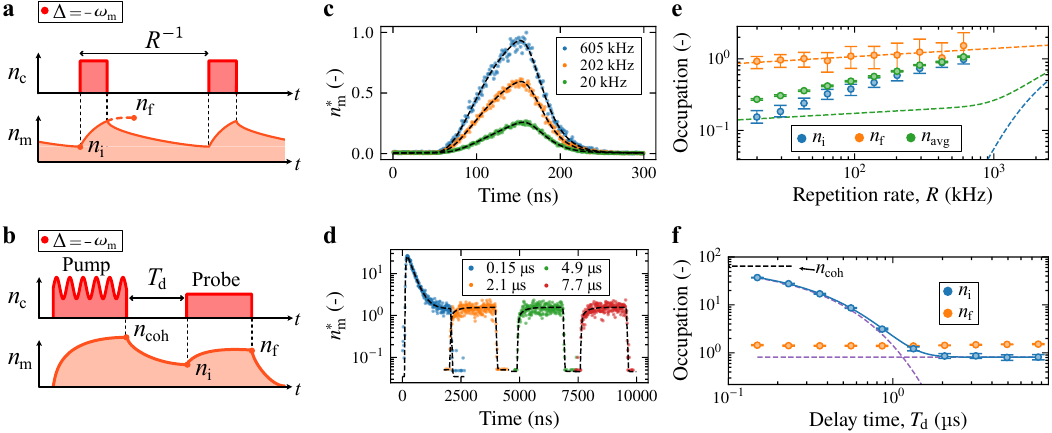}
    \caption{\textbf{Off-resonant pulsed noise dynamics of release-free OMC.} \textbf{a,c,e} Pulse train with 100 ns pulses at $\nc = 128$ for varying repetition rates ($R$). \textbf{b,d,f} Pump-probe experiment for varying probe delay times ($T_\mathrm{d}$) with repetition rate $R = \qty{70}{\kilo\hertz}$, \qty{2}{\micro\second} pulse durations and $\nc=48$. \textbf{a,b} Pulse schemes. $\nin$ ($\nf$) is the initial (final) occupation as defined in Eq.~\eqref{eq:thermal_noise_dynamics} whereas $n_\mathrm{coh}$ is the occupation reached on coherent driving. \textbf{c,d} Measured filtered mechanical occupation ($\nm^*(t)$) with fits to  Eq.~\eqref{eq:thermal_noise_dynamics} with applied filter transfer function. \textbf{e} Best fit parameters $[\nin,\nf]$ from \textbf{c} together with calculated average phonon occupation $n_\mathrm{avg}$  with 95\% credibility intervals (data points). We include lifetime-limited theory curves (dashed lines) showing the expected noise performance when $\nin$ decays exponentially in between pulses with time constant from Fig.~\ref{fig:pulse_exp}d. For this calculation we first fit $\nf(R)=(R/R_0)^\theta$ where $R_0 = \qty{52}{\kilo\hertz}$ and $\theta=0.114$. Next we calculate  $\nin(T_\mathrm{d}) = \nf\exp(-\gz T_\mathrm{d}) + n_0$ where $T_\mathrm{d} = R^{-1} - T_\mathrm{0}$ and $n_0$ is the dilution fridge occupation at $T_\mathrm{f}=\qty{50}{\milli\kelvin}$. Finally we get $n_\mathrm{avg}$ by averaging Eq.~\eqref{eq:thermal_noise_dynamics} with calculated values for $\nin,\nf$. 
    \textbf{f} Best fit parameters $[\nin,\nf]$ from \textbf{d} with 95\% credibility intervals (data points). We fit the initial occupation to $\nin(T_\mathrm{d}) = n_\mathrm{coh}\exp(-\gamma_\mathrm{coh}T_\mathrm{d}) + n_\mathrm{res}$ with $n_\mathrm{coh} = 64$, $\gamma_\mathrm{coh}/(2\pi)=\qty{607}{\kilo\hertz}$ and $n_\mathrm{res} = 0.81$ (blue line). The individual terms of $\nin(T_\mathrm{d})$ are shown as dashed purple lines.
    }
    \label{fig:short_pulse_noise}
\end{figure*}

We fit the resonant data for both devices to power laws $\nm(\nc) = \beta(\nc)^\alpha$, where we extract $\alpha = 0.29$ for the suspended device. This power-law exponent is consistent with the $\np(\nc)$ dependence previously reported for one-dimensional OMCs, although the proportionality constant $\beta$ is roughly 2.5 times smaller in this work \cite{maccabe_nano-acoustic_2020}. For the release-free device, one single power law is not sufficient to describe the resonant data which could be due to its non-trivial noise bath coupling (Fig.~\ref{fig:pulse_exp}c). We split up the power law fit into high and low power regions where $\alpha = 0.61,\ (0.44)$ for high (low) power regions respectively. Both the low- and high-power exponents measured for the release-free device exceed the $\alpha \approx 0.33$ measured for $\np(\nc)$ in suspended devices \cite{maccabe_nano-acoustic_2020,sonar_high-efficiency_2025}.

In contrast to the exponent for $\gp$, an elevated exponent for $\np(\nc)$ ($\alpha$-value) deviates from the prediction of a three-dimensional hot-bath model, which predicts $\alpha = 1/(d+1)$ for an effective bath dimensionality $d$ \cite{maccabe_nano-acoustic_2020}. Therefore, a higher $\alpha$-value in the region where $\gp$ dominates suggests a hot bath exhibiting lower-dimensionality behavior. Deviations of this kind occur also in related quantum devices, where coupling to localized two-level systems (TLS) can alter the phonon dynamics \cite{cleland_studying_2024, chen_phonon_2024}. A complete characterization of such TLS interactions lies beyond the scope of this work. However, we suspect that the etched silicon surfaces and the amorphous SiO$_2$ of our devices host a reservoir of TLS that can mediate thermal noise between the optical pump and the mechanical mode. 

Regarding the noise magnitude, we see a significantly lower thermal occupation for the release-free device for all detunings when compared to the suspended counterpart. At a thermal occupation near $\nm=1,$ the release-free device withstands \qty{29}{\decibel} more intracavity optical energy at off-resonant pumping compared to the suspended device. For resonant pumping, this increased optical power handling increases to \qty{35}{\decibel}. We attribute the increased noise for off-resonant pumping at equal $\nc$ to the bus waveguide providing heat to the OMC through the substrate. We expect the bus waveguide heating to be more prominent for off-resonant pumping due to the higher powers required to reach a certain $\nc$ \cite{ren_two-dimensional_2020,sonar_high-efficiency_2025}. The suspended device on the other hand has a less direct thermal link between bus waveguide and OMC which we observe as resonant and off-resonant data converging for low $\nc$. We therefore expect even lower off-resonant noise for release-free devices in the absence of bus waveguide heating effects. Despite the higher thermal noise for off-resonant pumping, the optomechanical back-action cools the mechanics for $\nc>10^3$ (Fig.~\ref{fig:CW_n_vs_nc}).

For a resonantly pumped cavity, the steady-state mechanical occupation follows from detailed balance as $\nm = (\gp\np + \gz n_0)/(\gz+\gp)$, where $n_0$ is the occupation of the cold phonon bath (Fig.~\ref{fig:release-free concept}d) \cite{ren_two-dimensional_2020}. For OMCs where $\gp \gg \gz$, we have $\nm \approx \np$ if assuming a cold dilution fridge bath. 
As mentioned above, this does not apply to the release-free device due to the stronger coupling to the cold bath $\gz$. The resonant release-free $\gm(\nc)$ reveals that $\gp<\gz$ for $\nc < 3500$ (Fig.~\ref{fig:pulse_exp}c). 
Therefore, the mechanical mode is predominantly coupled through $\gz$ to the cold bath at low $\nc$. 
Nevertheless, the resonant occupation levels measured for low $\nc$ (Fig.~\ref{fig:CW_n_vs_nc}) corresponds to a bath temperature of $T_\mathrm{f}>\qty{200}{\milli\kelvin}$ which far exceeds the steady state dilution fridge temperatures around 10 mK at low applied optical powers. 
This appears to be inconsistent with the thermal environment model (Fig.~\ref{fig:release-free concept}d) under the assumption of a cold bath properly thermalized with the \qty{10}{\milli\kelvin} cryostat. One hypothesis is that the immediate environment of the OMC -- representing the cold bath -- is at a higher temperature than the dilution fridge due to insufficient thermal transport away from the device. Another hypothesis is that a higher thermal occupation at low $\nc$ might still come from the hot bath if $\gp(\nc)$ includes a low exponent term that would dominate at low $\nc$ -- which is not excluded by our data (Fig.~\ref{fig:pulse_exp}c). Finally, there may be additional noise sources not currently included in the thermal model (Fig.~\ref{fig:release-free concept}d), such as the TLS ensembles suggested by the lower-dimensionality behavior of the noise scaling at high $\nc$ (Fig.~\ref{fig:CW_n_vs_nc}).

\subsection*{Noise occupancy in pulsed operation} \label{sec:pulsed noise performance}
We now return to pulsed measurements to characterize noise occupancy for off-resonant pumping. With detuning $\Delta = \pm\om$ an experiment can be configured to activate beam-splitter or two-mode-squeezing interactions, both relevant for quantum protocols \cite{aspelmeyer_cavity_2014,duan_long-distance_2001,krastanov_optically_2021}. In the following, we choose to pump red-detuned to avoid vacuum fluctuations, focusing on the thermal noise aspect. For the experiments in this section we focus on the release-free device that exhibits a mechanical decay rate $\gz$ compatible with fast repetition rate experiments.

We first run a short pulse $(T_\mathrm{0} = \qty{100}{\nano\second})$ experiment with power $(\nc=128)$ to reach an anti-Stokes scattering rate $p_\mathrm{as} = \gaom T_\mathrm{0} = 0.05$ (Fig.~\ref{fig:short_pulse_noise}a,c,e). These short and low-scattering-rate pulses are of interest for e.g. heralding experiments \cite{haug_heralding_2024, jiang_optically_2023,meesala_non-classical_2024,duan_long-distance_2001,krastanov_optically_2021}.
We proceed as in our previous pulsed experiments
to gather data and fit for $[\nin,\nf]$ with which we here calculate the average phonon occupancy during a pulse as $n_\mathrm{avg}=T_\mathrm{0}^{-1}\int \nm(t)dt$. By repeating the experiment for different delay times $T_\mathrm{d}$, we study the noise in each pulse as a function of the experiment's repetition rate $R = (T_\mathrm{0} + T_\mathrm{d})^{-1}$.

The noise dynamics observed in this experiment differ markedly from those in our previous decay measurements (Fig.~\ref{fig:pulse_exp}d). Contrary to the expected outcome, the initial occupation $(\nin)$ behaves non-exponentially across the explored repetition rates. In addition, the estimated final occupation $(\nf)$ also scales with repetition rate, suggesting a thermal build-up effect connected to the average dissipated optical power. With this measured performance, the release-free device can be operated under present experimental condition with $n_\mathrm{avg}<0.5$ at $R=\qty{100}{\kilo\hertz}$, limited by the slow decay of the initial population. This repetition rate ($R$-value) exceeds the decay rate of state-of-the-art superconducting qubits operating at \qty{5}{\giga\hertz} \cite{kono_mechanically_2024}. When simulating $n_\mathrm{avg}$ under the assumption of an exponentially-decaying $\nin$ with the intrinsic dynamics measured in Fig.~\ref{fig:pulse_exp}d, we estimate that the release-free device could perform quantum experiments with $n_\mathrm{avg}<0.3$ for repetition rates exceeding $R = \qty{1}{\mega\hertz}$.

Non-exponential noise dynamics as a function of repetition rate as seen in Fig.~\ref{fig:short_pulse_noise}e have also been reported for two-dimensional suspended OMCs \cite{mayor_high_2025}, although its origin remains unclear. To study this in more detail, we perform a second red-detuned experiment with a pump-probe pulse scheme (Fig.~\ref{fig:short_pulse_noise}b,d,f). The pump is chosen to be a \qty{2}{\micro\second} red-detuned pulse with amplitude modulation at the mechanical frequency. The modulated pump drives the mechanical mode into a coherent state with coherent occupation $n_\mathrm{coh}$ which decays freely after the end of the pump pulse. After a variable decay time $T_\mathrm{d}$, we read out the mechanical state with a \qty{2}{\micro\second} probe pulse. For all delay times, the repetition rate of the experiment is fixed to $R = \qty{70}{\kilo\hertz}$ and both pulses have optical cavity occupancy $\nc=48$. 

From the extracted initial occupations, we estimate that the pump pulse drives the mechanical mode into a coherent state with $n_\mathrm{coh} = 64$, after which it decays exponentially with a time constant consistent with that obtained in our resonant pulse measurements (Fig.~\ref{fig:pulse_exp}d). For delays $T_\mathrm{d} > \qty{2}{\micro\second}$, however, the decay saturates at $\nm \approx 0.8$. This saturation level depends on the experiment's repetition rate (App.~\ref{app:slow decaying hot bat}) and follows a trend similar to that of the short-pulse occupations (Fig.~\ref{fig:short_pulse_noise}e). We therefore conclude that the mechanical mode decays freely until reaching a limit set by a slowly decaying reservoir, whose population depends on the average optical power dissipated during the experiment. The decay of this reservoir determines the noise occupancy observed in Fig.~\ref{fig:short_pulse_noise}e. Although the origin and dynamics of this reservoir are not yet fully understood, it may be linked to the elevated continuous-wave noise (high $T_\mathrm{f}$) observed at low $\nc$ (Fig.~\ref{fig:CW_n_vs_nc}), suggesting a non-exponential decaying cold bath
. Finally, we note that this bath remains spatially confined, as the mixing-chamber temperature never exceeds 20 mK during the measurements in this section.

\section*{Discussion} \label{sec:discussion}

In this first-generation device, the release-free OMC withstands nearly two orders of magnitude higher optical power before the onset of thermo-optic nonlinearities, and it maintains near-unity phonon occupancy at three orders of magnitude higher optical power than suspended nanobeams. Together, these features make the release-free OMC a promising platform that enables high resilience to optical power without significantly limiting optomechanical coupling.
The release-free architecture exhibits a higher intrinsic mechanical loss rate $\gz$ than shielded suspended nanobeams. This reduces the single-photon cooperativity but simultaneously increases the device bandwidth and potentially also strengthens thermalization to the cryostat cold bath
(Fig.~\ref{fig:release-free concept}d). The relevance of $\gz$ is therefore application-dependent, and the present regime ($\gz/(2\pi)\sim\SI{100}{\kilo\hertz}$) is well-suited to near-term quantum experiments \cite{chen_low-noise_2026}. 

In time-resolved measurements, the release-free device exhibits a rapid initial rethermalization after optical excitation, with dynamics governed by the intrinsic mechanical decay time. However, we also observe a non-exponential decay for longer time scales, possibly arising from coupling to localized two-level systems, limited chip-to-cryostat heat transport or residual ITR. This slow decay limits the noise occupancy and currently prevents this first-generation release-free device to match state-of-the-art suspended 1D and 2D OMCs \cite{fiaschi_optomechanical_2021,mayor_high_2025,sonar_high-efficiency_2025} in this regard.  
 
Looking ahead, improvements in surface passivation and chip-to-cryostat thermalization could have impact on thermal dynamics and lead to reduced pulsed noise \cite{borselli_surface_2007}, while statistical analysis of the noise scaling can give insight into potential effects of fabrication-disorder on thermalization \cite{sonar_high-efficiency_2025}. The release-free OMC thus provide an appealing platform for investigating noise dynamics in greater depth. Regarding optomechanical performance, decreasing the optical mode volume -- currently exceeding that of the mechanical mode \cite{burger_design_2025,kolvik_clamped_2023} -- would enhance the single-photon optomechanical coupling strength and allow for lower pump powers at fixed interaction rates. 

Transferring the architecture to a sapphire substrate is a natural next step for release-free OMCs \cite{burger_design_2025}. Sapphire would further improve thermal conductivity \cite{bon-mardion_sub-kelvin_2026}, enable proper removal of amorphous oxides that may host noise-contributing TLS ensembles \cite{xie_two-dimensional_2026}, and facilitate direct integration with high-quality superconducting microwave circuits through capacitive \cite{zhao_quantum-enabled_2025} or piezoelectric \cite{meesala_non-classical_2024,jiang_optically_2023} elements in a fully release-free electro-optomechanical platform \cite{burger_design_2025}. Taken together, these advances position release-free optomechanics toward harnessing GHz phonons as coherent, low-noise intermediaries for emerging technologies in high-power optics and quantum communication, sensing, and computation.

\begin{acknowledgments}
We thank W. Wieczorek, N. J. Engelsen, P. Delsing, O. J. Painter, A. H. Safavi-Naeini and W. Jiang for helpful discussions. The devices were fabricated in the MyFab Nanofabrication Laboratory at Chalmers. We acknowledge support from the Knut and Alice Wallenberg foundation through the Wallenberg Centre for Quantum Technology through a WACQT Fellowship, from the European Research Council via Starting Grant 948265, and from the Swedish Foundation for Strategic Research via grant FFL21-0039. M. B. K. is supported by the Carlsberg Foundation, Internationalization Fellowship grant CF24-0448.
\end{acknowledgments}

\section*{Methods} \label{sec:methods}
\subsection*{Single-phonon counting} 
Throughout this study, we characterize the optically-induced thermal noise in our devices through single-phonon counting \cite{cohen_phonon_2015,meenehan_pulsed_2015}. This method involves the use of single-photon detectors (SPDs) to measure optomechanically scattered sideband photons as a gauge for the average mechanical occupation. When placing a pump laser in the vicinity of the OMC optical resonance, we selectively measure Stokes or anti-Stokes scattered photons by placing a pair of narrowband tunable Fabry--Pérot filters (\qty{13.6}{\mega\hertz} individual bandwidth) over the desired sideband (App.~\ref{app:optical_setup}). At dilution refrigerator temperatures, the gigahertz mechanical mode of our devices is inherently close to its motional ground state. In this regime, the mechanical vacuum fluctuations cause a measurable asymmetry in the amplitude of the optomechanically scattered sidebands \cite{safavi-naeini_laser_2013}. When the pump laser frequency $(\ola)$ is placed with a detuning $\Delta \coloneq \ola - \oo = \pm \om$ from the optical resonance frequency $(\oo)$, the rate of detected sideband photons is given by
\begin{equation}
    \label{eq:Gamma}
    \Gamma_\mathrm{\pm}(t) = \gaom(t)\eta_\mathrm{tot}\left(\nm(t) + \frac{1}{2}(1 \pm 1)\right) + \Gamma_\mathrm{noise}(t),
\end{equation}
for Stokes $(+)$ and anti-Stokes $(-)$ process' respectively. In Eq.~\eqref{eq:Gamma} we have that $\gaom(t) = 4\gom^2\nc(t)/\kappa$, $\eta_\mathrm{tot}$ is the total sideband photon detection efficiency and $\nm(t)$ is the average mechanical occupancy. The noise term $\Gamma_\mathrm{noise}(t)$ comprises both excess laser noise and external noise from dark counts and blackbody radiation (App.~\ref{app:nNEP}). Eq.~\eqref{eq:Gamma} provides a direct way to measure thermally excited phonons in both continuous-wave and pulsed measurements. When pumping resonantly, both processes are suppressed by the optical density of states causing a reduction in the detected photon rate by a factor $(\kappa/2\om)^2$.

\section*{Data availability} \label{sec:data_availability}
The data that support the findings of this study are openly available on Zenodo at \url{https://doi.org/10.5281/zenodo.21905753}. Additional data is available from the corresponding author upon reasonable request.

\section*{Author contributions} \label{sec:author_contributions}
J.K. fabricated devices, performed the experimental studies, analyzed data and wrote the manuscript. P.B. designed the devices and assisted with experimental logistics. D.H. contributed to the Bayesian methods used for optical pulse curve fitting. T.H.H. procured and installed the QM pulse sequencer hardware. J.F. assisted with experimental logistics. M.B.K. contributed to the coherent mechanical ring-up experiments and provided experimental guidance.  R.V.L. conceived the project, provided funding and supervised the work.

\appendix

\section{Fabrication}\label{app:fabrication}

For fabricating both release-free and suspended OMCs, we use SOI wafers (SOITEC, $\sigma>\qty{750}{\ohm\centi\meter}$) with \qty{220}{\nano\meter} device layer and \qty{2}{\micro\meter} buried oxide. We define the OMC pattern with ARP6200 resist and electron-beam lithography (Raith EBPG5200) and transfer it to the device layer with an HBr and Cl$_2$-based ICP-RIE etch. Following dry-etch, we clean the chips using a piranha solution (3:1, H$_2$SO$_4$:O$_2$H$_2$) and 2\% HF. Finally, the conventional nanobeam OMC chips are suspended in 5\% HF.

\begin{table*}
    \caption{Optomechanical properties at 10 mK for the studied devices.}
    \centering
    \begin{tabular}{|c||c|c|c|c|c|c|c|}
        \hline
        Device & $\oo/(2\pi)$ (THz) & $\om/(2\pi)$ (GHz)  & $\kappa/(2\pi)$ (GHz) & $\ke/2\pi$ (MHz) & $\gz/(2\pi)$ (kHz) & $\gom/(2\pi)$ (kHz) & Experiment\\
        \hline\hline
        Release-free & 194.5867 & 5.358 & 1.63 & 680 & 500 & 470 & All\\
        Suspended A &  195.6301 & 5.984 & 1.35 & 580 & - & 855 & Thermo-optics\\
        Suspended B &  195.7751 & 5.983 & 1.70 & 1030 & 0.5 & 1069 & Noise thermometry\\
         \hline
    \end{tabular}
    \label{tab:dev_params}
\end{table*}

\section{Experimental setup} \label{app:optical_setup}

\begin{figure*}
    \centering
    \includegraphics{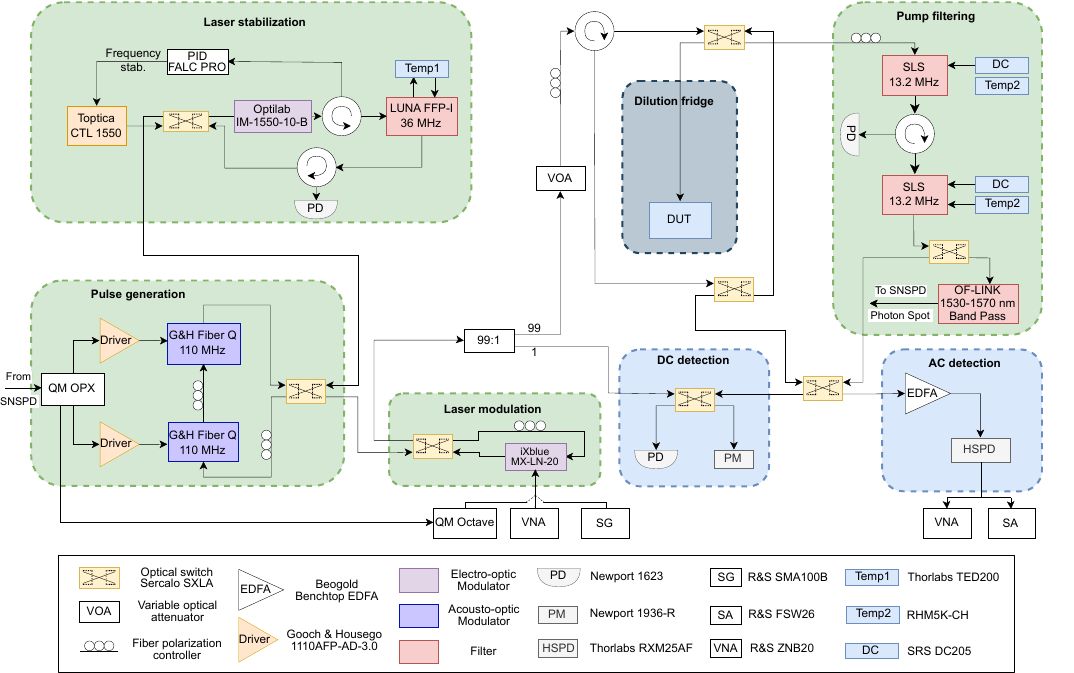}
    \caption{\textbf{Experimental setup.} We use a Toptica CTL 1550 to perform the studies presented in this work. We lock the laser frequency to a temperature tunable Fabry--Pérot filter with the Pound-Drever-Hall technique. This stabilization also serves to suppress laser phase noise at the mechanical frequency. Before sending the light to the device, we can apply acousto- and electro-optic modulation through in-path switching. The acousto-optic modulation is driven by a Quantum Machines OPX unit which handles sequencing of pulsing and single-photon detection. After preparing the laser light, we send it to the device under test (DUT) which is mounted to the mixing chamber of a dilution refrigerator. We choose to send the reflected light to either linear AC/CD detection using standard photodiodes, or to a single-photon counting path. Before detection, we use a set of two \qty{13.2}{\mega\hertz} wide tunable Fabry--Pérot filters to remove pump light and transmit the optomechanical sideband. Lastly, we suppress stray infrared radiation using a wide band pass filter before sending the light to superconducting nanowire single-photon detectors (SNSPD) operating at the \qty{900}{\milli\kelvin} stage of the cryostat.} 
    \label{fig:setup}
\end{figure*}

In this section, we describe the experimental setup used in this work (Fig.~\ref{fig:setup}). For the laser source, we use a tunable external cavity diode laser (Toptica CTL 1550). While the power of the laser is stabilized internally, we use optical switches to route the laser into a stabilization module for external frequency control. In the stabilization module, we use the Pound-Drever-Hall technique to lock the laser frequency to a fiber coupled Fabry-Pérot filter (LUNA FFP-I, bandwidth $=\qty{36}{\mega\hertz}$). The transmission frequency of the filter is tuned and stabilized through temperature control (Thorlabs TED200). This lock serves to stabilize the laser during long single-photon counting experiments in addition to suppressing phase noise and spontaneous emission amplitude noise at the mechanical frequency, crucial for accurate phonon thermometry \cite{safavi-naeini_laser_2013}.

After stabilization, we use optical switches to reroute the laser into laser modulators. Pulse modulation is achieved with a pair of 110 MHz acousto-optic modulators (G\&H Fiber Q), actuated through microwave drivers by a pulse sequencer module (QM OPX). With this setup, we synthesize pulses with 10-90\% rise time of \qty{14}{\nano\second} and extinction ratio $>\qty{100}{\decibel}$. Electro-optic amplitude modulation is used for EIT (App.~\ref{app:eit}), coherent mechanical driving (App.~\ref{app:calibration}) and pump filter alignment. For experiments with amplitude modulated pulses, we drive the EOM with a QM local oscillator unit controlled by the OPX. We send the pre-treated light through a circulator to the device under test placed in our dilution refrigerator system (Bluefors LD400). Before every cooldown, we pre-align to the OMC device using Attocube cryogenic piezo actuation stages. Alignment to the correct device is maintained throughout cryostat cooldown with an automated realignment scheme run every 30 minutes.

\begin{figure}
    \centering
    \includegraphics{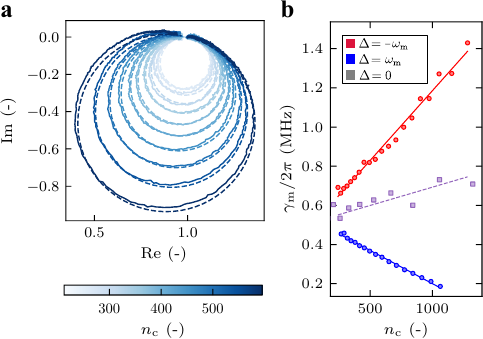}
    \caption{\textbf{Release-free electromagnetically induced transparency at 10 mK.} \textbf{a.} Normalized transparency window traces with fits to Eq.~\eqref{eq:EIT}. \textbf{b.} Total mechanical linewidth as function of $\nc$. Red and blue points are extracted from fits to the EIT data in \textbf{a}, while purple points are taken from Fig.~\ref{fig:pulse_exp}b of the main text. Red and blue solid lines are linear fits to the corresponding data sets revealing a slight asymmetry to the slopes. Averaging the linear fits yields the dashed purple line which follows the Fig.~\ref{fig:pulse_exp}b data with good agreement.}
    \label{fig:EIT}
\end{figure}

Depending on the current measurement, we direct the light reflected off the DUT to either linear or single-phonon detection paths. In the linear detection path, we use a photodetector (Newport 1623) and power meter (Newport 1936-R) for DC characterization and a high-speed photoreceiver (Thorlabs RXM25AF0) for AC sideband detection. In the single-photon detection path, the light passes through two fiber coupled filter cavities (Stable Laser Systems) to remove pump light. The Fabry--Pérot filters have \qty{13.2}{\mega\hertz} bandwidth and \qty{18.8}{\giga\hertz} FSR and provide \qty{114}{\decibel} pump suppression at \qty{5}{\giga\hertz} detuning when assembled in series. Next, we run the light through a broadband IR filter (OF-LINK Band Pass) to suppress unwanted photons outside the C-band before detection. We detect the transmitted sideband on superconducting nanowire singe-photon detectors (SNSPD) (Photon Spot) with output routed to the OPX for photon counting and pulse sequencing. 

Because of the very weak signals in the single-photon detection regime, active locking of the pump filters is not feasible. Instead, we use a reference sideband at the mechanical frequency, generated by our EOM, to align the filters to the correct detuning. Filter alignment during measurements is maintained through passive stabilization, combining acoustic and thermal isolation with resistive heater–based temperature control. For longer measurement runs, the filter stack is automatically realigned at intervals of 1–10 minutes. To suppress noise counts on the SNSPDs, the fiber linking the pump filter output to the dilution-fridge input is optically isolated by wrapping it in aluminum foil, and measurements are performed under minimal ambient light. 

\section{Device properties}\label{app:dev prop}

A full list of device properties is presented in Tab.~\ref{tab:dev_params}. All properties are measured at an ambient temperature of around \qty{10}{\milli\kelvin}. The study includes one release-free and two suspended OMCs with identical fabrication processes besides the final release step for the suspended devices. The two suspended devices are featured in different parts of the main text data. The thermo-optic effect was studied with suspended device A whereas the single-phonon-counting experiments were performed on suspended device B.

\section{Electromagnetically induced transparency}\label{app:eit}

Due to the low mechanical occupancy in the GHz mechanical modes at millikelvin temperatures the thermomechanical signature on the optical field is weak compared to room temperature. Using thermally-scattered sidebands for cryogenic study of optomechanical effects therefore requires high SNR measurements techniques such as balanced heterodyne detection \cite{meenehan_silicon_2014,ren_two-dimensional_2020} or single photon detection \cite{meenehan_pulsed_2015,maccabe_nano-acoustic_2020,ren_two-dimensional_2020,jiang_optically_2023, riedinger_remote_2018}. On the contrary, \textit{Electromagnetically Induced Transparency} (EIT) \cite{safavi-naeini_electromagnetically_2011} exploits coherently driven mechanical excitations for signal generation, thus making it independent of thermal population. In addition, OMC optical stability is often enhanced at low temperatures due to the suppressed thermo-optic effect. This makes EIT appealing for optomechanical back-action characterization at millikelvin temperatures.

We perform EIT measurements by placing an optical pump at blue or red motional sidebands of the OMC optical resonance. A weak sideband is then generated through electro-optic amplitude modulation driven by a vector network analyzer (VNA). We scan the modulation frequency while monitoring the optical reflection on a highspeed photodetector. A transparency feature around the mechanical frequency is measured with the VNA and can be described by a Lorentzian in the complex plane \cite{chan_laser_2012}
\begin{equation}
    \label{eq:EIT}
    S(\delta\omega) = 1 - \frac{A}{i\delta\omega - \gm/2}.
\end{equation}
In Eq.~\eqref{eq:EIT}, we have assumed a modulation frequency $\omega_\mathrm{mod} = \om + \delta\omega$ where $ \delta\omega \ll \kappa$, and use $A \in \mathbb{C}$ as a fitting parameter. In addition, Eq.~\eqref{eq:EIT} is obtained in the resolved-sideband regime and when normalizing to the response when $\delta\omega \gg \gm$.

We measure VNA traces for the release-free device taken at 10 mK when pumping on the blue motional sideband (Fig.~\ref{fig:EIT}a). Fitting the data with Eq.~\eqref{eq:EIT} yields back-action curves for both detunings. We notice a slight asymmetry in the linear slopes for red and blue sides causing disagreement when using the data to calculate optomechanical coupling. We remove the effect of optomechanical back-action by averaging the fitted red and blue data, revealing an additional power-dependent component to the linewidth (purple dashed line in Fig.~\ref{fig:EIT}b). Such asymmetric back-action curves could be explained by a Fano-shape to the optical resonance. However, we find that this power dependence agrees well with the linewidth data extracted from resonant pulsing (Fig.~\ref{fig:pulse_exp}b). Similar back-action asymmetry caused by thermal mechanical broadening has also been reported in lithium-niobate optomechanical crystals at low temperatures \cite{jiang_lithium_2019}.

\section{Thermometry calibration} \label{app:calibration}
Performing high-quality thermometry experiments requires precise and reliable calibration of the measured mechanical occupation. Using Eq.~\eqref{eq:Gamma} as a starting point, the problem of calibrating the phonon counting setup can be formulated as finding $\Gamma_\mathrm{cal}$ such that
\begin{equation}
    \label{eq:cal_rate}
    \nm = \frac{\Gamma_\mathrm{\pm}}{\Gamma_\mathrm{cal}\nc} - \frac{1}{2}(1\pm1),
\end{equation}
where $\Gamma_\mathrm{\pm}$ is the rate of detected Stokes $(+)$ anti-Stokes $(-)$ photons. $\Gamma_\mathrm{cal}$ can thus be thought of as being the anti-Stokes detection rate at unit intracavity photon and phonon numbers. In this work, we explore three methods of measuring $\Gamma_\mathrm{cal}$ based on different principles. In the following, we describe each method and assess its advantages and limitations.

\textit{Direct calibration} is the simplest of the three methods but requires full knowledge of OMC optomechanical properties and system detection efficiency. In the absence of noise, Eq.~\eqref{eq:Gamma} immediately yields 
\begin{equation}
    \label{eq:direct_cal}
    \Gamma_\mathrm{cal}^\mathrm{direct} = \frac{4\gom^2\eta_\mathrm{tot}}{\kappa},
\end{equation}
where we have assumed anti-Stokes scattering for simplicity. In our setup, the total efficiency can be broken up in $\eta_\mathrm{tot} = \eta_\mathrm{o}\eta_\mathrm{fc}\eta_\mathrm{loss}\eta_\mathrm{det}$, where $\eta_\mathrm{o} = \ke/\kappa$ is the optical coupling, $\eta_\mathrm{fc}$ is the fiber-chip coupling, $\eta_\mathrm{loss}$ represent photon loss from chip to detectors and $\eta_\mathrm{det}$ is the detection efficiency. To calculate this calibration rate, we start by measuring the photon loss from the mixing chamber to the SNSPD input at room temperature. Next, we measure the fiber-chip coupling when the cryostat is cold by comparing the device reflected light to the input. The detection efficiency of the SNSPDs is measured by sending light with known power to the detectors and measuring the recorded click rate. Finally, we measure the optomechanical device properties when the cryostat is cold (see main text). Even though simple in nature, the extensive prior system knowledge required leads us to use this calibration method as a cross-check for the following techniques.  

\begin{figure}
    \centering
    \includegraphics{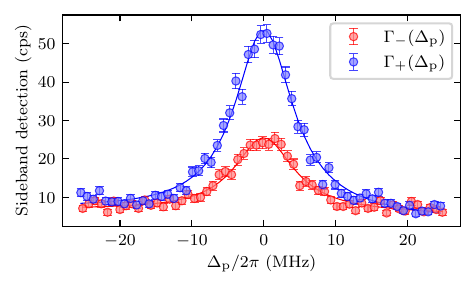}
    \caption{\textbf{Sideband asymmetry detection.} Stokes (blue) and anti-Stokes (blue) sidebands for the release-free device measured with single-photon detection. Error bars represent standard deviation assuming Poissonian statistics for photon arrival. We place the pump at $\Delta=\pm\om$ and measure the optomechanical sideband resonant with the OMC optical mode. By sweeping the resonance frequency of the pump filter stack, we trace out the sideband to find detection rates for thermometry calibration. Finally, we fit the data to Eq.~\eqref{eq:sideband_sweep}.}
    \label{fig:sideband_asym}
\end{figure}
\begin{figure*}
    \centering
    \includegraphics{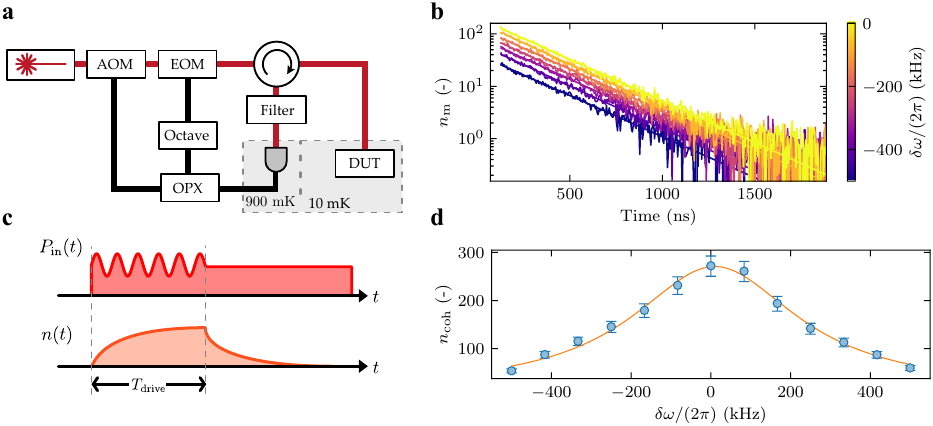}
    \caption{\textbf{Thermometry calibration through coherent excitation.} (a) Simplified experimental setup. A red detuned optical pulse train is generated using fast Acousto-Optic Modulation (AOM). Part of each pulse is modulated at the mechanical frequency with an amplitude Electro-Optic Modulator (EOM). Pump filters are used to selectively detect the anti-Stokes sideband using single photon detection. A pulse sequencer unit (QM OPX) orchestrates modulation pulsing. The MHz range idler output of the OPX is upconverted into the GHz regime with a local oscillator (QM Octave). (b) Experiment pulse scheme. We excite the OMC optical mode using a red detuned optical pulse train. During an initial time $T_\mathrm{drive}$, we drive the mechanical mode coherently by modulating the input at the mechanical frequency. After $T_\mathrm{drive}$, the drive modulation is turned off to detect the free mechanical decay. (c) Free mechanical decay with exponential fits after coherent driving at modulation frequency $\omega_\mathrm{mod} = \om + \delta\omega$. (d) Estimated phonon occupation $(n_\mathrm{coh})$ at the end of coherent drive ($t=0$ in (c)) for varying modulation frequencies with least squares standard deviation. Coherently excited phonons follow a Lorentzian around the mechanical frequency $\delta\omega = 0$.}
    \label{fig:coherent-excitation}
\end{figure*}

\textit{Sideband asymmetry.} At zero average mechanical occupation, the mechanical resonators ability to emit energy is completely suppressed; however, due to the vacuum fluctuations still present in the resonator, scattering events resulting in phonon absorption are still possible \cite{safavi-naeini_laser_2013}. This is the principle which governs the asymmetry in Stokes/anti-Stokes scattering rates at low phonon occupations as seen in Eq.~\eqref{eq:cal_rate}. We exploit this by measuring the photon detection rate for both Stokes $(\Gamma_\mathrm{+})$ and anti-Stokes $(\Gamma_\mathrm{-})$ scattering to calculate
\begin{equation}
    \Gamma_\mathrm{cal}^\mathrm{asym} = \frac{\Gamma_\mathrm{+} - \Gamma_\mathrm{-}}{\nc}.
    \label{eq:Gcal_asym}
\end{equation}
We measure the sidebands while pumping off-resonantly to achieve stronger signals at lower $\nm$. For both red and blue detuning, we use the pump filters to scan over the sideband resonant with the OMC optical mode. To accurately place the pump filters at an arbitrary detuning $\Delta_\mathrm{p}$, we use electro-optic modulation to drive a reference sideband at the desired detuning to which we align the filters. When the filters are correctly placed, we couple in the device and SNSPDs to measure $\Gamma_\pm(\Delta_\mathrm{p})$ (Fig.~\ref{fig:sideband_asym}). To take the pump filter transfer function into account we use that
\begin{equation}
    \Gamma_\pm(\Delta_\mathrm{p}) = \Gamma_\pm\int_{-\infty}^\infty T(\omega)^k\frac{\gm/(2\pi)d\omega}{(\omega - \Delta_\mathrm{p})^2 + (\gm/2)^2}.
    \label{eq:sideband_sweep}
\end{equation}
Here $k$ is the number of filters with transfer function $T(\omega) = (1 + (\omega/\kappa_\mathrm{pf})^2)^{-1}$ where $\kappa_\mathrm{pf}$ is the individual pump filter FWHM. For $\gm\ll\kappa_\mathrm{pf}$, Eq.~\eqref{eq:sideband_sweep} simplifies to $\Gamma_\pm(\Delta_\mathrm{p}) \approx \Gamma_\pm\cdot T(\Delta_\mathrm{p})^k$.

Due to vacuum fluctuations providing a reference, sideband asymmetry calibration requires less prior knowledge about the optomechanical system. This extra reference makes it an attractive choice when preparing for a thermometry experiment. Crucially, knowledge of the setup detection efficiency $\eta_\mathrm{tot}$ allows for calculation of the optomechanical coupling through Eq.~\eqref{eq:direct_cal}. This method can then be cross-referenced with other options for measuring coupling such as EIT. However, the sideband-asymmetry method also comes with complications that have to be taken into account. Firstly, excess laser noise at the mechanical frequency can cause both heating and noise squashing/anti-squashing \cite{kippenberg_phase_2013,safavi-naeini_laser_2013}. In addition, the method assumes equal experimental conditions when reading out both sidebands. When using one laser to read out both sidebands, care has to be taken to guarantee equal $\nc$ and $\nm$ for both red and blue detuned experiments. Therefore the power of the experiment must be low enough to ensure $\gaom\ll\gz+\gp$ to not cause excessive optomechanical back-action. Lastly, any error arising from noise or unequal experiment conditions for the two detunings propagates unfavorably when the asymmetry is weak, i.e. at insufficiently low $\nm$. Therefore, a proper sideband asymmetry calibration favors devices that achieve low $\nm$ for high pump powers to achieve higher signal-to-noise ratio. We also note that when using this calibration for pulsed experiments, we perform the calibration with CW excitation first and use the methods described in App.~\ref{app:filter_dynamics} to extract time-dependent average phonon occupations. Due to the time dynamics of the pump filters, we are not aware of a straightforward sideband-asymmetry calibration method for pulsed experiments when pulse durations are similar to the pump filter rise time.   

\textit{Coherent excitation.} This alternative to sideband asymmetry calibration also relies on a mechanical occupation reference; although this time, the reference is a coherent state with known amplitude. 
To drive the mechanics into a coherent state of known amplitude, we send an amplitude-modulated optical pulse train to the device. The amplitude modulation is driven at the mechanical frequency such that the optical beat note between pump and modulated sideband coherently drives the mechanics. After the mechanics has been driven during an initial time $T_\mathrm{drive}$ the modulation is terminated and the anti-Stokes sideband is read out using pump filters and single-photon detectors. During this time, the mechanical mode decays from the elevated coherent state back to the thermal state set by optical absorption. 

To calculate the expected coherent amplitude we follow the derivation given in Ref.~\cite{kristensen_long-lived_2024} to obtain
\begin{equation}
    \label{eq:n_coh}
    n_\mathrm{coh} = \frac{4\eta_\mathrm{o}\gamma_\mathrm{om}\xi_\mathrm{sb}\Phi_\mathrm{pump}}{\gamma_\mathrm{m}^2}.
\end{equation}
In this equation, $\xi_\mathrm{sb}$ is the modulated sideband power relative to the pump and $\Phi_\mathrm{pump}$ is the pump photon flux in the on-chip bus waveguide. In addition, we assume that the drive pulse is significantly longer than the response time of the mechanics $(\gm^{-1} \ll T_\mathrm{drive})$. The anti-Stokes detection rate measured at $\nm = n_\mathrm{coh}$ $(\Gamma_-^\mathrm{coh})$ now gives the calibration rate
\begin{equation}
    \Gamma_\mathrm{cal}^\mathrm{coh} = \frac{\Gamma_-^\mathrm{coh}}{\nc n_\mathrm{coh}}.
    \label{eq:Gcal_coherent}
\end{equation}

We perform the calibration for the release-free device with $T_\mathrm{drive} = \qty{2}{\micro\second}$ long drive pulses and study the mechanical decay for varying drive frequencies (Fig.~\ref{fig:coherent-excitation}c). As expected, the decay follows an exponential function with a time constant consistent with data presented in Fig.~\ref{fig:pulse_exp}d. Due to interference between the anti-Stokes scattered and electro-optic sidebands (EIT), we obtain the detection rate corresponding to the estimated $n_\mathrm{coh}$ through extrapolation. Using the interference-free part of the exponential decay (Fig.~\ref{fig:coherent-excitation}c) we estimate the detection rate at the end of the driving pulse ($t=0$).   

In summary, the various calibration techniques presented in this section provide both advantages and drawbacks. Due to being significantly more involved, we use coherent excitation calibration sparingly as a cross-check for the less time-consuming options. We therefore calibrate all data presented in the main text of this work with sideband asymmetry. However, we confirm that all calibration methods agree to within the uncertainty set by the ingoing parameters. We present example data of calibration rates measured through the three methods in Tab.~\ref{tab:cal_rates}.
\begin{table}
    \caption{\label{tab:cal_rates} Results from performing three different methods of calibration for the release-free device. We calculate the uncertainties by propagating estimated error intervals on all ingoing parameters used in Eqs.~\eqref{eq:direct_cal}, \eqref{eq:Gcal_asym}, \eqref{eq:Gcal_coherent}.}
    \centering
    \begin{tabular}{|c|c|}
        \hline
        Method & $\Gamma_\mathrm{cal}$ (cps) \\
        \hline
        \hline
        \textit{Direct} & $24.7\pm3.5$\\
        \textit{Sideband asymmetry} &  $21.1\pm7.0$ \\
        \textit{Coherent excitation} &  $20.8\pm2.1$ \\
         \hline
    \end{tabular}
\end{table}

\section{Pump filter dynamics in pulsed thermometry} \label{app:filter_dynamics}
The thermometry experiments in this work are mostly based on the phonon-counting technique described in the Method section of the main text. In this technique, we use tunable narrowband optical filters to extract the desired optomechanical sideband for detection on a single-photon detector. In our setup, the filter stack comprises a pair of $\qty{13.2}{\mega\hertz}$ bandwidth, $\qty{18.8}{\giga\hertz}$ FSR fiber coupled Fabry-Pérot filters (Stable Laser Systems). The filters are arranged in series and provide $<\qty{2}{\decibel}$ insertion loss at maximum transmission and \qty{114}{\decibel} pump suppression at \qty{5}{\giga\hertz} detuning from resonance. Though essential for the phonon counting measurement technique, the filter stack's bandwidth must be carefully considered when analyzing measurement results. Continuous-wave considerations are well covered in Ref.~\cite{sonar_high-efficiency_2025}; however, additional temporal dynamics must also be considered for experiments with short optical pulses.

We base the analysis on the amplitude transfer function for a lossless Fabry--Pérot filter comprising two identical mirrors with reflectivity $r$ given by Yu et al.~\cite{yu_optical_2001} as
\begin{equation}
    \label{eq:filter_transfer}
    t(\omega) = \frac{1-r}{1-r\exp(-i2\omega t_0)}e^{-i\omega t_0},
\end{equation}
where $t_0 = (2\cdot\mathrm{FSR})^{-1}$ is the optical passage time through the filter. We find the mirror reflectivity $r$ through the relation
\begin{equation}
    F = \frac{\pi\sqrt{r}}{1 - r},
\end{equation}
where $F$ is the filter finesse. This holds for high finesse filters which is the case for our filter stack where $F\approx1400$. We express the optical input field waveform as $E_\mathrm{in}(t) = \sqrt{P_\mathrm{in}(t)}\exp(-i\omega_\mathrm{sb}t)$, where $\omega_\mathrm{sb}$ is the sideband frequency and $P_\mathrm{in}(t)$ is the normalized input power envelope. Now, using Eq.~\eqref{eq:filter_transfer} we get the filter field output as
\begin{equation}
    E_\text{out}(t) = \mathcal{F}^{-1}\Bigl[\mathcal{F}[E_\text{in}]\cdot t(\omega)^k\Bigr],
    \label{eq:E_out}
\end{equation}
where $\mathcal{F}[\cdot]$ is the Fourier transform and $k$ is the number of identical filters in the stack. To get the normalized power output of the filter stack, we take the absolute square of Eq.~\eqref{eq:E_out} which simplifies when the input signal is resonant with a filter mode (e.g. $\omega_\text{sb} = \pi/t_0$). A change of variables in the Fourier transform yields
\begin{figure}
    \centering
    \includegraphics{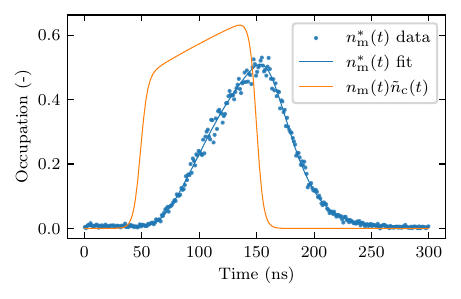}
    \caption{\textbf{Filter dynamics compensation.} We use the transfer function of the pump filters to derive the pulse-gated thermal occupation $\nm(t)\tilde{n}_\mathrm{c}(t)$ given calibrated filter output data $\nm^*(t)$ measured through single-photon detection. We take data points from Fig.~\ref{fig:short_pulse_noise}c of the main text, representing the anti-Stokes sideband detected when exciting the release-free device with $T_0 = \qty{100}{\nano\second}$ pulses detuned to $\Delta = -\om$. From the fit, we extract $\nin=0.42$ and $\nf=1.25$ at a repetition rate of \qty{33}{\kilo\hertz}.}
    \label{fig:filter_dynamics}
\end{figure}
\begin{equation}
    P_\text{out}(t) = \Bigl|\mathcal{F}^{-1}\Bigl[\mathcal{F}\Bigl[\sqrt{P_\text{in}}\Bigr]\cdot t(\omega)^k\Bigr]\Bigr|^2.
    \label{eq:P_out}
\end{equation}
Next, we use Eqs.~\eqref{eq:Gamma} and \eqref{eq:direct_cal} to connect the dynamics of the filter input optical power to that of the optomechanical occupations $\nm,\nc$. When omitting $\Gamma_\mathrm{noise}$ we get
\begin{equation}
    P_\text{in}(t) = \Gamma_\mathrm{cal}\nc^\mathrm{max}\hbar\oo(\nm(t)+(1\pm1)/2)\tilde{n}_\mathrm{c}(t),
\end{equation}
where $\tilde{n}_\mathrm{c}(t) = \nc(t)/\nc^\mathrm{max}$ is the normalized input pulse power. Finally, defining the calibrated SPD signal as $\nm^*(t) = P_\mathrm{out}(t)/(\Gamma_\mathrm{cal}\nc^\mathrm{max}\hbar\oo)$ we obtain
\begin{equation}
    \begin{split}
       \nm^*(t) = \\
       &\Bigl|\mathcal{F}^{-1}\Bigl[\mathcal{F}\Bigl[\sqrt{(\nm+(1\pm 1)/2)\tilde{n}_\mathrm{c}}\Bigr]\cdot t(\omega)^k\Bigr]\Bigr|^2.
        \label{eq:n_m^star}
    \end{split}
\end{equation}

In summary, with access to calibrated SPD data $\nm^*(t)$, AOM pulse envelop $\tilde{n}_\mathrm{c}(t)$ and filter transfer function $t(\omega)$, we use Eq.~\eqref{eq:n_m^star} to infer the phonon occupation dynamics $\nm(t)$ and its parameters through Eq.~\eqref{eq:thermal_noise_dynamics}. We calibrate the AOM waveform $\tilde{n}_\mathrm{c}(t)$ by measuring the output of the AOMs directly on our SPD with heavy attenuation. 

\begin{figure*}
    \centering
    \includegraphics{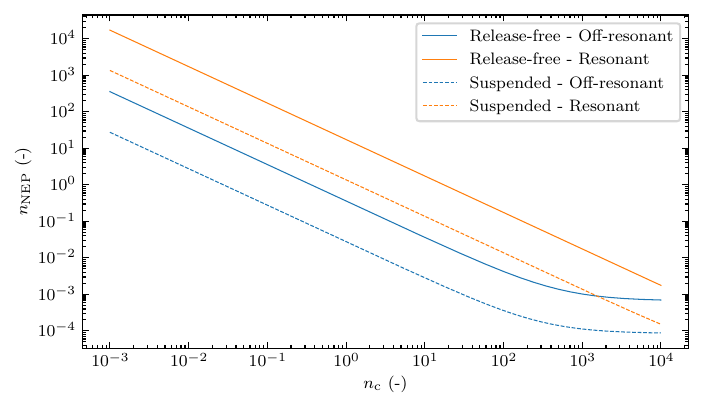}
    \caption{\textbf{Noise equivalent phonon occupation}. We plot the noise equivalent phonon occupation as a function of optical power given in units of $\nc$.}
    \label{fig:nNEP}
\end{figure*}

In Fig.~\ref{fig:filter_dynamics}, we show an example fit to \qty{100}{\nano\second} pulsed anti-Stokes sideband data with \qty{33}{\kilo\hertz} repetition rate (Fig.~\ref{fig:short_pulse_noise}a,c,e). In this fit, we use a premeasured $\gm$ as input parameter and fit the data for $[\nin,\nf,t_\text{stop},n_\mathrm{NEP}]$. The parameter $t_\text{stop}$ represents the input pulse stop time and $n_\mathrm{NEP}$ is the phonon occupation equivalent with the detector dark counts, entering as a constant addition to Eq.~\eqref{eq:n_m^star}. Note that by using Eq.~\eqref{eq:thermal_noise_dynamics} for $\nm(t)$, we omit any dynamics in the thermal drive caused by the finite rise-time of the optical pulse. For a more detailed description on our fitting see App.~\ref{app:bayesian_fitting}.  

\section{Bayesian fitting methods} \label{app:bayesian_fitting}

Inferring the dynamics of the mechanical mode during optical pulsing is made complicated when the timescale of the filter dynamics is comparable
to the dynamics of the mechanical mode occupation. This is generally the case for our release-free device where $\gm/\kappa_\mathrm{pf}>0.01$ with $\kappa_\mathrm{pf}$ as the FWHM of an individual pump filter. A further complication is that the low photon scattering rate at low powers leads to low amounts of data. Due to the above, the range of model parameter values that are consistent with measured data may be large when the filter transfer functions obfuscate the data. In this work, we use Bayesian
parameter estimation and Markov chain Monte Carlo (MCMC) methods to both fit the parameters of our model and, crucially,
get an accurate estimate for the uncertainty in our estimation.

Given a model $M$ that takes model parameters $\theta$, collected data $D$, and prior information $I$, we seek to infer the posterior probability distribution $p(\theta | D, M, I)$. This probability can be rewritten using Bayes rule as 
\begin{equation}
    \label{eq:bayes}
    p(\theta | D, M) = \frac{p(D | \theta, M) p(\theta | M)}{p(D|M)}, 
\end{equation}
where we omit $I$ for brevity. 

In Eq.~\eqref{eq:bayes}, $p(D | \theta, M)$ is the likelihood which model the probability of measuring the data $D$ given certain model parameters $\theta$. 
In our case, the models for phonon and filter dynamics along with system losses give a predicted photon detection rate during the extent of a pulse $\Gamma_\theta(t)$ (see Eq.~\eqref{eq:Gamma} and App.~\ref{app:filter_dynamics}). During an SPD measurement, we collect clicks into time bins of size $\Delta t$. When assuming Poissonian statistics for incoming photon clicks, the probability of measuring $k_i$ clicks in the time bin $[t_i, t_i+\Delta t]$ is
\begin{equation}
    p_{i}(k_i) = \frac{\lambda_i^{k_i} e^{-\lambda}}{k_i!},
\end{equation}
where $\lambda_i = \Gamma_\theta(t_i) \Delta t$. Therefore, we get the likelihood as the probability of measuring the data $D$, consisting of counts $k_1, k_2, \ldots, k_N$ in time bins $t_1, t_2 \ldots, t_N$ through
\begin{equation}
    p(D | \theta, M) = \prod_{i=1}^N p_i.
\end{equation}
Next, the factor $p(\theta | M)$ in Eq.~\eqref{eq:bayes} encapsulates the prior beliefs on ingoing parameters which we choose as uniform distributions over sensible values. Finally, we consider the evidence of the model, $p(D | M)$, as an inconsequential normalization constant due to its independence on $\theta$.

To analyze the posterior distribution, we use MCMC methods, specifically the \texttt{emcee} python package \cite{foreman-mackey_emcee_2013}, which optimizes the likelihood for $\theta$ and samples the posterior distribution $p(\theta|D,M)$. We get credibility intervals for each parameter in the model by taking appropriate quantiles of the marginalized distribution. This ensures that the uncertainty in any nuisance parameters are properly propagated to the uncertainty of the parameters of interest.

\section{Noise equivalent phonon occupation} \label{app:nNEP}
After proper thermometry calibration (App.~\ref{app:calibration}), we can estimate the \textit{noise equivalent phonon occupation} ($n_\mathrm{NEP}$) which serves as a soft lower limit to the measurable phonon occupation for any given experiment. We assume that the noise measured during SPD comes primarily from two sources; detection events in the absence of laser light (commonly referred to as dark counts) and excess laser noise. Dark counts include internally generated detector clicks, black-body radiation and stray light from e.g. laboratory lighting. The excess laser noise stems either from light at the pump frequency not perfectly attenuated by the pump filters or from laser noise process' at the mechanical frequency such as laser phase noise \cite{safavi-naeini_laser_2013}. 

We measure the excess laser noise by first aligning the pump filter stack to be detuned from the pump laser by the mechanical frequency. Next, we redirect the pump laser to bypass the device and directly detect its transmission through the pump filters. Without pre-filtering of the pump laser, we measure a photon flux on the SPD equivalent to \qty{-92}{\decibel c} before the pump filters. This noise is likely dominated by phase noise and spontaneous emission induced amplitude noise. When engaging pump pre-filtering, the measured noise photon flux drops to \qty{-113.6}{\decibel c}. This number is consistent with the \qty{114}{\decibel} expected pump suppression given by Eq.~\eqref{eq:filter_transfer} assuming two filters with FWHM = \qty{13.2}{\mega\hertz} and FSR = \qty{18.8}{\giga\hertz}.

\begin{figure}
    \centering
    \includegraphics{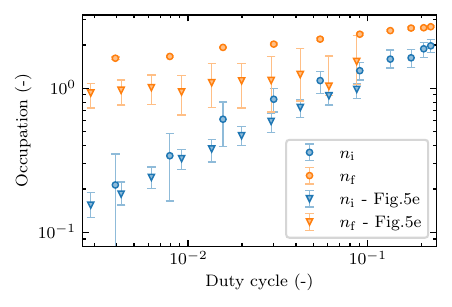}
    \caption{\textbf{Thermal noise limit duty cycle dependence.} Initial $(\nin)$ and final $(\nf)$ occupation data from pump-probe experiment without coherent driving. In this experiment, pump and probe pulse durations are \qty{2}{\micro\second},  $\nc = 324$ and $T_\mathrm{d} = \qty{10}{\micro\second}$. We vary the duty cycle of the experiment by controlling the delay between pulse pairs. In addition, we plot occupation data from Fig.~\ref{fig:short_pulse_noise}e. Data is presented with 95\% credibility intervals.}
    \label{fig:duty cycle experiment}
\end{figure}

\begin{figure*}
    \centering
    \includegraphics{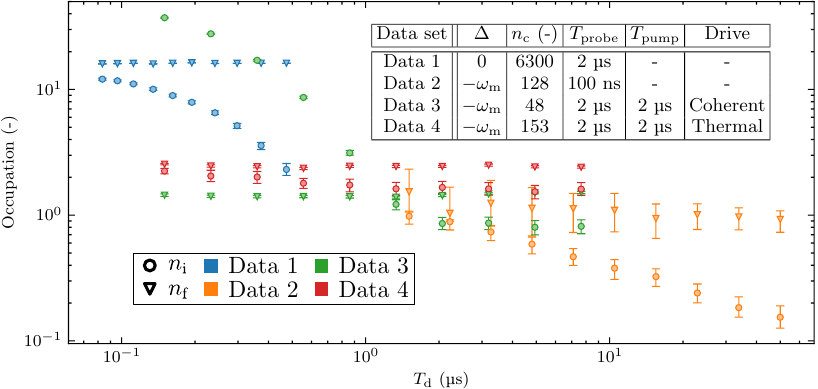}
    \caption{\textbf{Overview of release-free data.} Release-free mechanical occupation measured in the pulsed experiments throughout this study. We take Data 1 from  Fig.~\ref{fig:pulse_exp}d, Data 2 from Fig.~\ref{fig:short_pulse_noise}e, and Data 3 from Fig.~\ref{fig:short_pulse_noise}f. Data 4 represents a pump-probe $T_\mathrm{d}$ sweep with no coherent drive during the pump pulse. For Data 1 and 2, $T_\mathrm{d}$ refers to the downtime between subsequent pulses, whilst it refers to the delay between pump and probe for Data 3 and 4.}
    \label{fig:master_ringdown}
\end{figure*}

Now, the noise equivalent phonon occupation is given by the ratio between expected noise photons and the calibration rate at given $\nc$. This ratio differs depending on the detuning and is given by \cite{meenehan_pulsed_2015}
\begin{align}
    &n_\mathrm{NEP}|_{\Delta = \pm\om} = \frac{\Gamma_\mathrm{DCR}}{\Gamma_\mathrm{cal}\nc} + A\left(\frac{\kappa\om}{2\ke\gom}\right)^2,\\
    &n_\mathrm{NEP}|_{\Delta = 0} = \frac{\Gamma_\mathrm{DCR}}{\Gamma_\mathrm{cal}\nc}\left(\frac{2\om}{\kappa}\right)^2 + A\left(\frac{\kappa\om(1-2\eta_\mathrm{o})}{\ke\gom}\right)^2,
\end{align}
where $\Gamma_\mathrm{DCR}$ is the detected dark count rate and $A$ is the pump suppression. We plot the expected $n_\mathrm{NEP}$ for both release-free and suspended devices in Fig.~\ref{fig:nNEP}. In the experiments with our release-free device, we are mainly limited by dark count noise $\Gamma_\mathrm{DCR}\approx\qty{7}{\hertz}$ and fiber-chip coupling $\eta_\mathrm{fc}\approx0.1$. We expect that with reasonable setup improvements, $\Gamma_\mathrm{cal}$ can be improved by more than 5 dB giving the same improvement in $n_\mathrm{NEP}$.

\section{Slowly decaying reservoir dynamics} \label{app:slow decaying hot bat}
In the the main text we study the ringdown of a coherently excited mechanical occupation and find that after an initial time governed by intrinsic mechanical decay ($\gz$), a noise floor at $\nm\approx0.8$ limits further drop in thermal noise for $T_\mathrm{d}>\qty{2}{\micro\second}$.(Fig.~\ref{fig:short_pulse_noise}f). In that experiment the repetition rate was fixed at $R=\qty{70}{\kilo\hertz}$. In this section, we show that this measured noise floor can be decreased by decreasing the average optical power supplied to the device.

We now run another pump-probe experiment with \qty{2}{\micro\second} pulse durations, although this time we remove the EOM driving to exclude the possibility of the noise floor originating from EOM noise at the mechanical frequency. The power of the experiment is set to be $\nc = 324$ and $T_\mathrm{d} = \qty{10}{\micro\second}$ while the delay between pulse pairs is varied, effectively controlling the experiment's repetition rate. We proceed to fit the data of the probe pulse and extract $\nin$ and $\nf$ as function of the duty cycle of the experiment; i.e., the fractional experiment time when optical power is on, including both pump and probe pulses. We see a clear reduction in both $\nin$ and $\nf$ for lower duty cycle pulse trains (Fig.~\ref{fig:duty cycle experiment}). In addition, we see a duty-cycle dependence in the ratio $\nin/\nf$, suggesting that the reservoirs decay dynamics and coupling to the mechanical mode also depend on average dissipated optical power.

The observed duty-cycle dependence of initial and final occupations is reminiscent of the short pulse noise in Fig.5e. By plotting the data from Fig.~\ref{fig:short_pulse_noise}e together with the data from this pump-probe experiment we see good agreement in the behavior of $\nin$ as a function of duty cycle (Fig.~\ref{fig:duty cycle experiment}). We thus conclude that the thermal noise decay measured in Fig.~\ref{fig:short_pulse_noise}e is limited by the dynamics of the measured background in Fig.~\ref{fig:short_pulse_noise}f. As pulsed noise performance is critical for application within near-term quantum experiments, it is of high importance to understand the source of this noise in future work. We conjecture that identifying this source is not only of interest for the release-free architecture, but also for other silicon optomechanical platforms operating with high repetition rate and low added noise.

\section{Compiled release-free ringdown} \label{app:master_ringdown}
For ease of overview, we report on all pulsed ring-down experiments performed throughout this work (Fig.~\ref{fig:master_ringdown}). We also present a thermally-driven pump-probe experiment to check that the noise floor measured in Fig.~\ref{fig:short_pulse_noise}f is not caused by residual EOM noise at the mechanical frequency. We perform the experiment identically to Fig.~\ref{fig:short_pulse_noise}b except having the EOM switched out of the laser path and increasing the power to $\nc=153$. For this experiment, we immediately observe the slow decay dynamics in $\nin$ from $T_\mathrm{d}>\qty{2}{\micro\second}$, ruling out the EOM as the source of the noise floor in Fig.~\ref{fig:short_pulse_noise}f.

\section{Application in a piezo-optomechanical quantum transducer} \label{app:transducer_efficiency}
\subsection*{Efficiency-bandwidth product}

An important application for optomechanical crystals is as key components for quantum-enabled microwave-optical transducers \cite{zou_microwave-optical_2021}. Functionalizing the mechanical mode of the optomechanical crystal with an electromechanically active region (through e.g. piezoelectricity) has seen much success in recent years towards this goal \cite{jiang_optically_2023,meesala_non-classical_2024,meesala_quantum_2024,weaver_integrated_2024}. In this section, we discuss the release-free device's potential when integrated into such a transducer system.

We note that the results in this section are based on projections using reasonable parameter values obtained from simulations \cite{burger_design_2025} and from separate measurements \cite{meesala_non-classical_2024}. A full experimental study is required to concretely settle the practical performance of a release-free transducer beyond projections and should be subject of future study.

\begin{figure}
    \centering
    \includegraphics{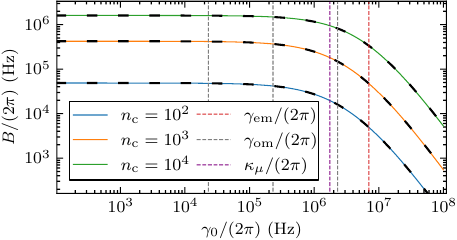}
    \caption{\textbf{Transducer efficiency-bandwidth product.} Simulated efficiency-bandwidth products $B = \int\eta(\omega)d\omega$ (Eq.~\eqref{eq:transducer_efficiency}) as function of intrinsic mechanical loss rate $\gz$ for different optical occupations $\nc$. Black dashed lines indicate a phenomenological model given in Eq.~\eqref{eq:EBP_model}. Vertical dashed lines represents electromechanical scattering rate $\gamma_{\rm em}$, optomechanical scattering rate $\gaom$ and microwave total loss rate $\kappa_\mu$. $\gaom = 4\gom\nc/\kappa_\mathrm{o}$ scales with $\nc$ and thus have one line per $\nc$-value. Auxiliary parameters are taken from \cite{burger_design_2025,meesala_non-classical_2024} and listed in Tab.~\ref{tab:transducer_params}. }
    \label{fig:efficiency_bandwidth}
\end{figure}

In a continuous-wave setting, the relevant figures of merit for the transducer include the conversion efficiency, conversion bandwidth and the added noise. Assuming that the transducer is coupled to a microwave resonator to enhance electromechanical coupling \cite{jiang_optically_2023,meesala_non-classical_2024,zhao_quantum-enabled_2025}, the total transducer efficiency can be calculated through coupled mode- and input-output theory \cite{shi_overcoming_2024} 
\begin{align}
    &\eta(\omega) = \frac{G_\mathrm{om}^2g_\mathrm{em}^2\kappa_\mathrm{\mu e}\kappa_\mathrm{o e}}{|G_\mathrm{om}^2\chi_\mu^{-1} + g_\mathrm{em}^2\chi_\mathrm{o}^{-1} + \chi_\mu^{-1} \chi_\mathrm{o}^{-1} \chi_\mathrm{m}^{-1}|^2}, \label{eq:transducer_efficiency}\\
    &\chi_\mu=\chi_\mu(\omega) = (-i(\omega - \omega_\mu) + \kappa_\mu/2)^{-1},\\
    &\chi_\mathrm{o}=\chi_\mathrm{o}(\omega) = (-i(\omega + \Delta) + \kappa_\mathrm{o}/2)^{-1},\\
    &\chi_\mathrm{m}=\chi_\mathrm{m}(\omega) = (-i(\omega - \omega_\mathrm{m}) + \gamma_\mathrm{i}/2)^{-1}.
\end{align}
In the above, we introduce the effective optomechanical coupling $G_\mathrm{om} = \gom\sqrt{\nc}$,  the electromechanical coupling $g_\mathrm{em}$ and internal mechanical losses $\gamma_\mathrm{i} = \gz+\gp$. We also define total and external loss rates for the optical mode $\kappa_\mathrm{o}, \kappa_\mathrm{oe}$ and for the microwave mode $\kappa_\mu, \kappa_{\mu\mathrm{e}}$, as well as the microwave mode frequency $\omega_\mu$.

Using Eq.~\eqref{eq:transducer_efficiency}, we evaluate the effective bandwidth of the transducer through the efficiency-bandwidth product $B = \int\eta(\omega)d\omega$ \cite{zhao_quantum-enabled_2025,shi_overcoming_2024}. We pull expected parameters for a release-free piezo-optomechanical transducer from \cite{burger_design_2025} coupled to a high-impedance microwave resonator \cite{meesala_non-classical_2024}. All simulation parameters are summarized in Tab.~\ref{tab:transducer_params}. In Fig.~\ref{fig:efficiency_bandwidth}, we plot the calculated efficiency-bandwidth product $B$ as a function of $\gamma_\mathrm{i}$ for a few different $\nc$. For the assumed parameters, $B$ is largely insensitive of $\gamma_\mathrm{i}$ up to a cut-off.

\begin{table}[]
    \centering
    \caption{Piezo-optomechanical parameters used for simulating frequency-dependent transducer efficiency (Eq.~\eqref{eq:transducer_efficiency}).}
    \begin{tabular}{c|r}
         \label{tab:transducer_params}
         Parameter& Value $(2\pi)$ \\
         \hline
         $\omega_\mathrm{m}$& \SI{4.77}{\giga\hertz} \\
         $\omega_\mu$& \SI{4.77}{\giga\hertz} \\
         $\Delta$& $-\SI{4.77}{\giga\hertz}$ \\
         $\kappa_\mathrm{o}$& \SI{1}{\giga\hertz} \\
         $\kappa_\mathrm{oe}$& \SI{0.5}{\giga\hertz} \\
         $\kappa_\mathrm{\mu}$& \SI{1.75}{\mega\hertz} \\
         $\kappa_\mathrm{\mu e}$& \SI{1.2}{\mega\hertz} \\
         $g_\mathrm{om}$& \SI{0.24}{\mega\hertz} \\
         $g_\mathrm{em}$& \SI{6}{\mega\hertz} \\
    \end{tabular}
    
    \label{tab:placeholder}
\end{table}

To better understand the efficincy-bandwidth product, we derive a phenomenological model for $B$ based on Eq.~\eqref{eq:transducer_efficiency} given by
\begin{equation}
    \label{eq:EBP_model}
    B\approx2\pi\eta_\mathrm{o}\gamma_\mathrm{om}\frac{\gamma_\mathrm{em}}{\gamma_\mathrm{em} + \gamma_\mathrm{om} + \gamma_\mathrm{i}}\frac{\kappa_\mathrm{\mu e}}{\kappa_\mu + \gamma_\mathrm{om} + \gamma_\mathrm{i}},
\end{equation}
in which $\gamma_\mathrm{em} = 4g_\mathrm{em}^2/\kappa_\mu$ is the electromechanical scattering rate. With the above model, we find good agreement for all simulated $\nc$ (Fig.~\ref{fig:efficiency_bandwidth}). From Eq.~\eqref{eq:EBP_model} we find that the transducer performance is not affected by mechanical losses while $\gamma_\mathrm{i} < \mathrm{max}(\gaom,\mathrm{min}(\gamma_\mathrm{em},\kappa_\mu))$ which relaxes constrains on acceptable mechanical losses. Intuitively, this can be thought of as the transducer being unaffected by intrinsic mechanical losses while there are stronger decay channels present for the mechanical mode. For the transducer parameters given in Tab.~\ref{tab:transducer_params} the cut-off is given by $\kappa_\mu/(2\pi) = \SI{1.75}{\mega\hertz}$ due to the strong electromechanical coupling.

\subsection*{Transducer added noise}

The transducer added noise in a continuous-wave setting can also be derived from the coupled-mode theory used to find Eq.~\eqref{eq:transducer_efficiency}. Following the thermal environment model introduced in Fig.~\ref{fig:release-free concept}d we describe the input-referred added noise levels for up- and downconversion as
\begin{align}
    n_\mathrm{add,up} = \frac{n_\mathrm{i}}{\eta_\mu C_\mathrm{em}},\\
    n_\mathrm{add,down} = \frac{n_\mathrm{i}}{\eta_\mathrm{o}C_\mathrm{om}},
\end{align}
where $C_\mathrm{om/em} = \gamma_\mathrm{om/em}/\gamma_\mathrm{i}$ are the opto-/electromechanical cooperativities and $\eta_\mathrm{\mu} = \kappa_\mathrm{\mu e}/\kappa_\mathrm{\mu }$ \cite{zou_microwave-optical_2021}. The parameter $n_\mathrm{i}$ represents the noise occupation in the baths connected to the mechanical mode's intrinsic decay rate $\gamma_\mathrm{i}$. When assuming uncorrelated noise in the hot and cold baths, the occupation $n_\mathrm{i}$ is given by $n_\mathrm{i} = (n_0\gz + \np\gp)/\gamma_\mathrm{i}$ \cite{zhao_quantum-enabled_2025}.

When assuming a majority noise phonon influx from the hot bath with rate $\gp\np\gg\gz n_0$, we see that the input-referred noise is insensitive to the cold bath coupling $\gz$. This model therefore predicts that the added noise in a transducer setting is solely dependent on hot-bath properties, electro-/optomechanical scattering rates and optical/microwave coupling efficiencies.

In summary, the coupling to the cryostat cold bath $\gz$ is but one of several interactions affecting the mechanical mode in a transducer setting. The efficiency–bandwidth product is sensitive to increases in $\gz$ only when this channel dominates the total decay rate. Assessing the influence of $\gz$ on added noise further requires an explicit model for phonon influx, such as the hot-bath model depicted in  Fig.~\ref{fig:release-free concept}d. When the cold bath is not the primary contributor to the mechanical mode’s thermal noise, variations in $\gz$ have a negligible impact on the input-referred added noise of the transducer. Critically, achieving both high efficiency and low noise necessitates a mechanical environment with a low effective temperature, together with large scattering rates $\gamma_\mathrm{em}$ and $\gamma_\mathrm{om}$ -- conditions that are well supported by the release-free platform.

\bibliographystyle{naturemag_noURL}
\bibliography{references}

@misc{xie_two-dimensional_2026,
	title = {A two-dimensional piezo-optomechanical transducer},
	url = {http://arxiv.org/abs/2607.26161},
	doi = {10.48550/arXiv.2607.26161},
	urldate = {2026-08-03},
	publisher = {arXiv},
	author = {Xie, Tian and Ooi, Nelson and Woodard, Linus and Malik, Sultan and Mayor, Felix and Hitchcock, Oliver A. and Primo, André G. and Jiang, Wentao and Gyger, Samuel and Safavi-Naeini, Amir H.},
	month = jul,
	year = {2026},
	note = {arXiv:2607.26161 [quant-ph]},
}

@article{meenehan_pulsed_2015,
	title = {Pulsed excitation dynamics of an optomechanical crystal resonator near its quantum ground state of motion},
	volume = {5},
	url = {https://link.aps.org/doi/10.1103/PhysRevX.5.041002},
	doi = {10.1103/PhysRevX.5.041002},
	number = {4},
	urldate = {2023-09-07},
	journal = {Physical Review X},
	publisher = {American Physical Society},
	author = {Meenehan, Seán M. and Cohen, Justin D. and MacCabe, Gregory S. and Marsili, Francesco and Shaw, Matthew D. and Painter, Oskar},
	month = oct,
	year = {2015},
	pages = {041002},
}

@article{meesala_quantum_2024,
	title = {Quantum entanglement between optical and microwave photonic qubits},
	volume = {14},
	url = {https://link.aps.org/doi/10.1103/PhysRevX.14.031055},
	doi = {10.1103/PhysRevX.14.031055},
	number = {3},
	urldate = {2025-06-12},
	journal = {Physical Review X},
	publisher = {American Physical Society},
	author = {Meesala, Srujan and Lake, David and Wood, Steven and Chiappina, Piero and Zhong, Changchun and Beyer, Andrew D. and Shaw, Matthew D. and Jiang, Liang and Painter, Oskar},
	month = sep,
	year = {2024},
	pages = {031055},
}

@article{qiu_laser_2020,
	title = {Laser cooling of a nanomechanical oscillator to {Its} zero-point energy},
	volume = {124},
	issn = {0031-9007, 1079-7114},
	url = {https://link.aps.org/doi/10.1103/PhysRevLett.124.173601},
	doi = {10.1103/PhysRevLett.124.173601},
	number = {17},
	urldate = {2025-10-16},
	journal = {Physical Review Letters},
	author = {Qiu, Liu and Shomroni, Itay and Seidler, Paul and Kippenberg, Tobias J.},
	month = apr,
	year = {2020},
	pages = {173601},
}

@article{chan_laser_2011,
	title = {Laser cooling of a nanomechanical oscillator into its quantum ground state},
	volume = {478},
	issn = {1476-4687},
	url = {https://www.nature.com/articles/nature10461},
	doi = {10.1038/nature10461},
	number = {7367},
	urldate = {2021-10-01},
	journal = {Nature},
	publisher = {Nature Publishing Group},
	author = {Chan, Jasper and Alegre, T. P. Mayer and Safavi-Naeini, Amir H. and Hill, Jeff T. and Krause, Alex and Gröblacher, Simon and Aspelmeyer, Markus and Painter, Oskar},
	month = oct,
	year = {2011},
	pages = {89--92},
}

@article{zou_microwave-optical_2021,
	title = {Microwave-optical quantum frequency conversion},
	volume = {8},
	issn = {2334-2536},
	url = {https://www.osapublishing.org/viewmedia.cfm?uri=optica-8-8-1050&seq=0&html=true},
	doi = {10.1364/OPTICA.425414},
	number = {8},
	urldate = {2021-09-01},
	journal = {Optica},
	publisher = {Optical Society of America},
	author = {Zou, Chang-Ling and Tang, Hong X. and Jiang, Liang and Fu, Wei and Han, Xu and Tang, Hong X. and Tang, Hong X.},
	month = aug,
	year = {2021},
	pages = {1050--1064},
}

@misc{bon-mardion_sub-kelvin_2026,
	title = {Sub-kelvin thermal conductivity of substrates and on-chip routing in quantum integrated systems},
	url = {http://arxiv.org/abs/2605.06146},
	doi = {10.48550/arXiv.2605.06146},
	urldate = {2026-07-25},
	publisher = {arXiv},
	author = {Bon-Mardion, Charles and Lorin, Arnaud and Deschaseaux, Edouard and Feautrier, Céline and Mermin, Daniel and Charbonnier, Jean and Li, Jing and Sauvageot, Jean-Luc and Thomas, Candice},
	month = may,
	year = {2026},
	note = {arXiv:2605.06146 [cond-mat.mes-hall]},
}

@article{jiang_lithium_2019,
	title = {Lithium niobate piezo-optomechanical crystals},
	volume = {6},
	copyright = {© 2019 Optical Society of America},
	issn = {2334-2536},
	url = {https://opg.optica.org/optica/abstract.cfm?uri=optica-6-7-845},
	doi = {10.1364/OPTICA.6.000845},
	number = {7},
	urldate = {2026-07-22},
	journal = {Optica},
	publisher = {Optica Publishing Group},
	author = {Jiang, Wentao and Patel, Rishi N. and Mayor, Felix M. and McKenna, Timothy P. and Arrangoiz-Arriola, Patricio and Sarabalis, Christopher J. and Witmer, Jeremy D. and Laer, Raphaël Van and Safavi-Naeini, Amir H.},
	month = jul,
	year = {2019},
	pages = {845--853},
}

@article{fiaschi_optomechanical_2021,
	title = {Optomechanical quantum teleportation},
	volume = {15},
	copyright = {2021 The Author(s), under exclusive licence to Springer Nature Limited},
	issn = {1749-4893},
	url = {https://www.nature.com/articles/s41566-021-00866-z},
	doi = {10.1038/s41566-021-00866-z},
	number = {11},
	urldate = {2026-01-28},
	journal = {Nature Photonics},
	publisher = {Nature Publishing Group},
	author = {Fiaschi, Niccolò and Hensen, Bas and Wallucks, Andreas and Benevides, Rodrigo and Li, Jie and Alegre, Thiago P. Mayer and Gröblacher, Simon},
	month = nov,
	year = {2021},
	pages = {817--821},
}

@article{weaver_integrated_2024,
	title = {An integrated microwave-to-optics interface for scalable quantum computing},
	volume = {19},
	copyright = {2023 The Author(s), under exclusive licence to Springer Nature Limited},
	issn = {1748-3395},
	url = {https://www.nature.com/articles/s41565-023-01515-y},
	doi = {10.1038/s41565-023-01515-y},
	number = {2},
	urldate = {2025-02-11},
	journal = {Nature Nanotechnology},
	publisher = {Nature Publishing Group},
	author = {Weaver, Matthew J. and Duivestein, Pim and Bernasconi, Alexandra C. and Scharmer, Selim and Lemang, Mathilde and van Thiel, Thierry C. and Hijazi, Frederick and Hensen, Bas and Gröblacher, Simon and Stockill, Robert},
	month = feb,
	year = {2024},
	pages = {166--172},
}

@article{jiang_optically_2023,
	title = {Optically heralded microwave photon addition},
	volume = {19},
	copyright = {2023 The Author(s), under exclusive licence to Springer Nature Limited},
	issn = {1745-2481},
	url = {https://www.nature.com/articles/s41567-023-02129-w},
	doi = {10.1038/s41567-023-02129-w},
	number = {10},
	urldate = {2025-12-01},
	journal = {Nature Physics},
	publisher = {Nature Publishing Group},
	author = {Jiang, Wentao and Mayor, Felix M. and Malik, Sultan and Van Laer, Raphaël and McKenna, Timothy P. and Patel, Rishi N. and Witmer, Jeremy D. and Safavi-Naeini, Amir H.},
	month = oct,
	year = {2023},
	pages = {1423--1428},
}

@article{krastanov_optically_2021,
	title = {Optically {Heralded} {Entanglement} of {Superconducting} {Systems} in {Quantum} {Networks}},
	volume = {127},
	url = {https://link.aps.org/doi/10.1103/PhysRevLett.127.040503},
	doi = {10.1103/PhysRevLett.127.040503},
	number = {4},
	urldate = {2026-03-17},
	journal = {Physical Review Letters},
	publisher = {American Physical Society},
	author = {Krastanov, Stefan and Raniwala, Hamza and Holzgrafe, Jeffrey and Jacobs, Kurt and Lončar, Marko and Reagor, Matthew J. and Englund, Dirk R.},
	month = jul,
	year = {2021},
	pages = {040503},
}

@article{shi_overcoming_2024,
	title = {Overcoming the fundamental limit of quantum transduction via intraband entanglement},
	volume = {2},
	copyright = {https://doi.org/10.1364/OA\_License\_v2\#VOR-OA},
	issn = {2837-6714},
	url = {https://opg.optica.org/abstract.cfm?URI=opticaq-2-6-475},
	doi = {10.1364/opticaq.540881},
	number = {6},
	urldate = {2025-07-11},
	journal = {Optica Quantum},
	publisher = {Optica Publishing Group},
	author = {Shi, Haowei and Zhuang, Quntao},
	month = dec,
	year = {2024},
	pages = {475},
}

@article{chen_low-noise_2026,
	title = {Low-noise optomechanical single phonon-photon conversion for quantum networks},
	volume = {17},
	copyright = {2026 The Author(s)},
	issn = {2041-1723},
	url = {https://www.nature.com/articles/s41467-025-67956-2},
	doi = {10.1038/s41467-025-67956-2},
	language = {en},
	number = {1},
	urldate = {2026-02-13},
	journal = {Nature Communications},
	publisher = {Nature Publishing Group},
	author = {Chen, Liu and Korsch, Alexander Rolf and Kersul, Cauê Moreno and Benevides, Rodrigo and Yu, Yong and Alegre, Thiago P. Mayer and Gröblacher, Simon},
	month = jan,
	year = {2026},
	pages = {1187},
}

@article{hong_hanbury_2017,
	title = {Hanbury {Brown} and {Twiss} interferometry of single phonons from an optomechanical resonator},
	volume = {358},
	url = {https://www.science.org/doi/10.1126/science.aan7939},
	doi = {10.1126/science.aan7939},
	number = {6360},
	urldate = {2026-02-06},
	journal = {Science},
	publisher = {American Association for the Advancement of Science},
	author = {Hong, Sungkun and Riedinger, Ralf and Marinković, Igor and Wallucks, Andreas and Hofer, Sebastian G. and Norte, Richard A. and Aspelmeyer, Markus and Gröblacher, Simon},
	month = oct,
	year = {2017},
	pages = {203--206},
}

@article{patel_engineering_2017,
	title = {Engineering {Phonon} {Leakage} in {Nanomechanical} {Resonators}},
	volume = {8},
	url = {https://link.aps.org/doi/10.1103/PhysRevApplied.8.041001},
	doi = {10.1103/PhysRevApplied.8.041001},
	number = {4},
	urldate = {2026-02-06},
	journal = {Physical Review Applied},
	publisher = {American Physical Society},
	author = {Patel, Rishi N. and Sarabalis, Christopher J. and Jiang, Wentao and Hill, Jeff T. and Safavi-Naeini, Amir H.},
	month = oct,
	year = {2017},
	pages = {041001},
}

@article{safavi-naeini_controlling_2019,
	title = {Controlling phonons and photons at the wavelength scale: integrated photonics meets integrated phononics},
	volume = {6},
	issn = {2334-2536},
	url = {https://www.osapublishing.org/viewmedia.cfm?uri=optica-6-2-213&seq=0&html=true},
	doi = {10.1364/OPTICA.6.000213},
	number = {2},
	urldate = {2021-09-01},
	journal = {Optica},
	publisher = {Optical Society of America},
	author = {Safavi-Naeini, Amir H. and Van Thourhout, Dries and Van Laer, Raphaël and Baets, Roel},
	month = feb,
	year = {2019},
	pages = {213--232},
}

@article{borselli_surface_2007,
	title = {Surface encapsulation for low-loss silicon photonics},
	volume = {91},
	issn = {0003-6951},
	url = {https://doi.org/10.1063/1.2793820},
	doi = {10.1063/1.2793820},
	number = {13},
	urldate = {2025-11-25},
	journal = {Applied Physics Letters},
	author = {Borselli, M. and Johnson, T. J. and Michael, C. P. and Henry, M. D. and Painter, O.},
	month = sep,
	year = {2007},
	pages = {131117},
}

@article{kolvik_clamped_2023,
	title = {Clamped and sideband-resolved silicon optomechanical crystals},
	volume = {10},
	copyright = {© 2023 Optica Publishing Group},
	issn = {2334-2536},
	url = {https://opg.optica.org/optica/abstract.cfm?uri=optica-10-7-913},
	doi = {10.1364/OPTICA.492143},
	number = {7},
	urldate = {2024-08-06},
	journal = {Optica},
	publisher = {Optica Publishing Group},
	author = {Kolvik, Johan and Burger, Paul and Frey, Joey and Van Laer, Raphaël},
	month = jul,
	year = {2023},
	pages = {913--916},
}

@article{cleland_studying_2024,
	title = {Studying phonon coherence with a quantum sensor},
	volume = {15},
	copyright = {2024 The Author(s)},
	issn = {2041-1723},
	url = {https://www.nature.com/articles/s41467-024-48306-0},
	doi = {10.1038/s41467-024-48306-0},
	number = {1},
	urldate = {2025-10-13},
	journal = {Nature Communications},
	publisher = {Nature Publishing Group},
	author = {Cleland, Agnetta Y. and Wollack, E. Alex and Safavi-Naeini, Amir H.},
	month = jun,
	year = {2024},
	pages = {4979},
}

@article{zhao_quantum-enabled_2025,
	title = {Quantum-enabled microwave-to-optical transduction via silicon nanomechanics},
	volume = {20},
	copyright = {2025 The Author(s), under exclusive licence to Springer Nature Limited},
	issn = {1748-3395},
	url = {https://www.nature.com/articles/s41565-025-01874-8},
	doi = {10.1038/s41565-025-01874-8},
	number = {5},
	urldate = {2025-10-16},
	journal = {Nature Nanotechnology},
	publisher = {Nature Publishing Group},
	author = {Zhao, Han and Chen, William David and Kejriwal, Abhishek and Mirhosseini, Mohammad},
	month = may,
	year = {2025},
	pages = {602--608},
}

@article{ren_two-dimensional_2020,
	title = {Two-dimensional optomechanical crystal cavity with high quantum cooperativity},
	volume = {11},
	issn = {2041-1723},
	url = {https://www.nature.com/articles/s41467-020-17182-9},
	doi = {10.1038/s41467-020-17182-9},
	number = {1},
	urldate = {2022-09-25},
	journal = {Nature Communications 2020 11:1},
	publisher = {Nature Publishing Group},
	author = {Ren, Hengjiang and Matheny, Matthew H. and MacCabe, Gregory S. and Luo, Jie and Pfeifer, Hannes and Mirhosseini, Mohammad and Painter, Oskar},
	month = jul,
	year = {2020},
	pages = {1--10},
}

@article{duan_long-distance_2001,
	title = {Long-distance quantum communication with atomic ensembles and linear optics},
	volume = {414},
	copyright = {2001 Macmillan Magazines Ltd.},
	issn = {1476-4687},
	url = {https://www.nature.com/articles/35106500},
	doi = {10.1038/35106500},
	number = {6862},
	urldate = {2025-10-16},
	journal = {Nature},
	publisher = {Nature Publishing Group},
	author = {Duan, L.-M. and Lukin, M. D. and Cirac, J. I. and Zoller, P.},
	month = nov,
	year = {2001},
	pages = {413--418},
}

@article{sonar_high-efficiency_2025,
	title = {High-efficiency low-noise optomechanical crystal photon-phonon transducers},
	volume = {12},
	copyright = {© 2025 Optica Publishing Group},
	issn = {2334-2536},
	url = {https://opg.optica.org/optica/abstract.cfm?uri=optica-12-1-99},
	doi = {10.1364/OPTICA.538557},
	number = {1},
	urldate = {2025-02-24},
	journal = {Optica},
	publisher = {Optica Publishing Group},
	author = {Sonar, Sameer and Hatipoglu, Utku and Meesala, Srujan and Lake, David P. and Ren, Hengjiang and Painter, Oskar},
	month = jan,
	year = {2025},
	pages = {99--104},
}

@article{kono_mechanically_2024,
	title = {Mechanically induced correlated errors on superconducting qubits with relaxation times exceeding 0.4 ms},
	volume = {15},
	copyright = {2024 The Author(s)},
	issn = {2041-1723},
	url = {https://www.nature.com/articles/s41467-024-48230-3},
	doi = {10.1038/s41467-024-48230-3},
	number = {1},
	urldate = {2025-10-16},
	journal = {Nature Communications},
	publisher = {Nature Publishing Group},
	author = {Kono, Shingo and Pan, Jiahe and Chegnizadeh, Mahdi and Wang, Xuxin and Youssefi, Amir and Scigliuzzo, Marco and Kippenberg, Tobias J.},
	month = may,
	year = {2024},
	pages = {3950},
}

@article{doeleman_brillouin_2023,
	title = {Brillouin optomechanics in the quantum ground state},
	volume = {5},
	issn = {2643-1564},
	url = {https://link.aps.org/doi/10.1103/PhysRevResearch.5.043140},
	doi = {10.1103/PhysRevResearch.5.043140},
	number = {4},
	urldate = {2025-10-16},
	journal = {Physical Review Research},
	author = {Doeleman, H. M. and Schatteburg, T. and Benevides, R. and Vollenweider, S. and Macri, D. and Chu, Y.},
	month = nov,
	year = {2023},
	pages = {043140},
}

@article{barclay_nonlinear_2005,
	title = {Nonlinear response of silicon photonic crystal microresonators excited via an integrated waveguide and fiber taper},
	volume = {13},
	copyright = {© 2005 Optical Society of America},
	issn = {1094-4087},
	url = {https://opg.optica.org/oe/abstract.cfm?uri=oe-13-3-801},
	doi = {10.1364/OPEX.13.000801},
	number = {3},
	urldate = {2024-10-25},
	journal = {Optics Express},
	publisher = {Optica Publishing Group},
	author = {Barclay, Paul E. and Srinivasan, Kartik and Painter, Oskar},
	month = feb,
	year = {2005},
	pages = {801--820},
}

@article{kippenberg_phase_2013,
	title = {Phase noise measurement of external cavity diode lasers and implications for optomechanical sideband cooling of {GHz} mechanical modes},
	volume = {15},
	issn = {1367-2630},
	url = {https://dx.doi.org/10.1088/1367-2630/15/1/015019},
	doi = {10.1088/1367-2630/15/1/015019},
	number = {1},
	urldate = {2025-07-23},
	journal = {New Journal of Physics},
	publisher = {IOP Publishing},
	author = {Kippenberg, T J and Schliesser, A and Gorodetsky, M L},
	month = jan,
	year = {2013},
	pages = {015019},
}

@article{foreman-mackey_emcee_2013,
	title = {emcee: {The} {MCMC} {Hammer}},
	volume = {125},
	issn = {1538-3873},
	shorttitle = {emcee},
	url = {https://iopscience.iop.org/article/10.1086/670067/meta},
	doi = {10.1086/670067},
	number = {925},
	urldate = {2025-09-18},
	journal = {Publications of the Astronomical Society of the Pacific},
	publisher = {IOP Publishing},
	author = {Foreman-Mackey, Daniel and Hogg, David W. and Lang, Dustin and Goodman, Jonathan},
	month = feb,
	year = {2013},
	pages = {306},
}

@article{yu_optical_2001,
	title = {Optical pulse propagation in a {Fabry}–{Perot} etalon: analytical discussion},
	volume = {18},
	copyright = {https://doi.org/10.1364/OA\_License\_v1\#VOR},
	issn = {1084-7529, 1520-8532},
	shorttitle = {Optical pulse propagation in a {Fabry}–{Perot} etalon},
	url = {https://opg.optica.org/abstract.cfm?URI=josaa-18-9-2153},
	doi = {10.1364/JOSAA.18.002153},
	number = {9},
	urldate = {2025-06-19},
	journal = {Journal of the Optical Society of America A},
	author = {Yu, Jin and Yuan, Shi and Gao, Jin-Yue and Sun, Lianzhi},
	month = sep,
	year = {2001},
	pages = {2153},
}

@article{safavi-naeini_laser_2013,
	title = {Laser noise in cavity-optomechanical cooling and thermometry},
	volume = {15},
	issn = {1367-2630},
	url = {https://dx.doi.org/10.1088/1367-2630/15/3/035007},
	doi = {10.1088/1367-2630/15/3/035007},
	number = {3},
	urldate = {2024-12-03},
	journal = {New Journal of Physics},
	publisher = {IOP Publishing},
	author = {Safavi-Naeini, Amir H. and Chan, Jasper and Hill, Jeff T. and Gröblacher, Simon and Miao, Haixing and Chen, Yanbei and Aspelmeyer, Markus and Painter, Oskar},
	month = mar,
	year = {2013},
	pages = {035007},
}

@article{cohen_phonon_2015,
	title = {Phonon counting and intensity interferometry of a nanomechanical resonator},
	volume = {520},
	copyright = {2015 Springer Nature Limited},
	issn = {1476-4687},
	url = {https://www.nature.com/articles/nature14349},
	doi = {10.1038/nature14349},
	number = {7548},
	urldate = {2025-06-27},
	journal = {Nature},
	publisher = {Nature Publishing Group},
	author = {Cohen, Justin D. and Meenehan, Seán M. and MacCabe, Gregory S. and Gröblacher, Simon and Safavi-Naeini, Amir H. and Marsili, Francesco and Shaw, Matthew D. and Painter, Oskar},
	month = apr,
	year = {2015},
	pages = {522--525},
}

@article{safavi-naeini_electromagnetically_2011,
	title = {Electromagnetically induced transparency and slow light with optomechanics},
	volume = {472},
	copyright = {2011 Nature Publishing Group, a division of Macmillan Publishers Limited. All Rights Reserved.},
	issn = {1476-4687},
	url = {https://www.nature.com/articles/nature09933},
	doi = {10.1038/nature09933},
	number = {7341},
	urldate = {2023-03-10},
	journal = {Nature},
	publisher = {Nature Publishing Group},
	author = {Safavi-Naeini, A. H. and Alegre, T. P. Mayer and Chan, J. and Eichenfield, M. and Winger, M. and Lin, Q. and Hill, J. T. and Chang, D. E. and Painter, O.},
	month = apr,
	year = {2011},
	note = {Number: 7341},
	pages = {69--73},
}

@article{huang_room-temperature_2024,
	title = {Room-temperature quantum optomechanics using an ultralow noise cavity},
	volume = {626},
	copyright = {2024 The Author(s)},
	issn = {1476-4687},
	url = {https://www.nature.com/articles/s41586-023-06997-3},
	doi = {10.1038/s41586-023-06997-3},
	number = {7999},
	urldate = {2025-10-13},
	journal = {Nature},
	publisher = {Nature Publishing Group},
	author = {Huang, Guanhao and Beccari, Alberto and Engelsen, Nils J. and Kippenberg, Tobias J.},
	month = feb,
	year = {2024},
	pages = {512--516},
}

@article{riedinger_non-classical_2016,
	title = {Non-classical correlations between single photons and phonons from a mechanical oscillator},
	volume = {530},
	copyright = {2016 Springer Nature Limited},
	issn = {1476-4687},
	url = {https://www.nature.com/articles/nature16536},
	doi = {10.1038/nature16536},
	number = {7590},
	urldate = {2025-04-16},
	journal = {Nature},
	publisher = {Nature Publishing Group},
	author = {Riedinger, Ralf and Hong, Sungkun and Norte, Richard A. and Slater, Joshua A. and Shang, Juying and Krause, Alexander G. and Anant, Vikas and Aspelmeyer, Markus and Gröblacher, Simon},
	month = feb,
	year = {2016},
	pages = {313--316},
}

@article{riedinger_remote_2018,
	title = {Remote quantum entanglement between two micromechanical oscillators},
	volume = {556},
	copyright = {2018 Macmillan Publishers Ltd., part of Springer Nature},
	issn = {1476-4687},
	url = {https://www.nature.com/articles/s41586-018-0036-z},
	doi = {10.1038/s41586-018-0036-z},
	number = {7702},
	urldate = {2024-09-27},
	journal = {Nature},
	publisher = {Nature Publishing Group},
	author = {Riedinger, Ralf and Wallucks, Andreas and Marinković, Igor and Löschnauer, Clemens and Aspelmeyer, Markus and Hong, Sungkun and Gröblacher, Simon},
	month = apr,
	year = {2018},
	pages = {473--477},
}

@article{mayor_high_2025,
	title = {High photon-phonon pair generation rate in a two-dimensional optomechanical crystal},
	volume = {16},
	copyright = {2025 The Author(s)},
	issn = {2041-1723},
	url = {https://www.nature.com/articles/s41467-025-57948-7},
	doi = {10.1038/s41467-025-57948-7},
	number = {1},
	urldate = {2025-10-14},
	journal = {Nature Communications},
	publisher = {Nature Publishing Group},
	author = {Mayor, Felix M. and Malik, Sultan and Primo, André G. and Gyger, Samuel and Jiang, Wentao and Alegre, Thiago P. M. and Safavi-Naeini, Amir H.},
	month = mar,
	year = {2025},
	pages = {2576},
}

@article{meesala_non-classical_2024,
	title = {Non-classical microwave–optical photon pair generation with a chip-scale transducer},
	volume = {20},
	copyright = {2024 The Author(s), under exclusive licence to Springer Nature Limited},
	issn = {1745-2481},
	url = {https://www.nature.com/articles/s41567-024-02409-z},
	doi = {10.1038/s41567-024-02409-z},
	number = {5},
	urldate = {2025-06-12},
	journal = {Nature Physics},
	publisher = {Nature Publishing Group},
	author = {Meesala, Srujan and Wood, Steven and Lake, David and Chiappina, Piero and Zhong, Changchun and Beyer, Andrew D. and Shaw, Matthew D. and Jiang, Liang and Painter, Oskar},
	month = may,
	year = {2024},
	pages = {871--877},
}

@article{chen_phonon_2024,
	title = {Phonon engineering of atomic-scale defects in superconducting quantum circuits},
	volume = {10},
	url = {https://www.science.org/doi/10.1126/sciadv.ado6240},
	doi = {10.1126/sciadv.ado6240},
	number = {37},
	urldate = {2025-09-18},
	journal = {Science Advances},
	publisher = {American Association for the Advancement of Science},
	author = {Chen, Mo and Owens, John Clai and Putterman, Harald and Schäfer, Max and Painter, Oskar},
	month = sep,
	year = {2024},
	pages = {eado6240},
}

@article{lagasse_higherorder_1973,
	title = {Higher‐order finite‐element analysis of topographic guides supporting elastic surface waves},
	volume = {53},
	issn = {0001-4966},
	url = {https://doi.org/10.1121/1.1913432},
	doi = {10.1121/1.1913432},
	number = {4},
	urldate = {2025-09-08},
	journal = {The Journal of the Acoustical Society of America},
	author = {Lagasse, P. E.},
	month = apr,
	year = {1973},
	pages = {1116--1122},
}

@article{haug_heralding_2024,
	title = {Heralding entangled optical photons from a microwave quantum processor},
	volume = {22},
	url = {https://link.aps.org/doi/10.1103/PhysRevApplied.22.034068},
	doi = {10.1103/PhysRevApplied.22.034068},
	number = {3},
	urldate = {2025-07-02},
	journal = {Physical Review Applied},
	publisher = {American Physical Society},
	author = {Haug, Trond Hjerpekjøn and Kockum, Anton Frisk and Van Laer, Raphaël},
	month = sep,
	year = {2024},
	pages = {034068},
}

@phdthesis{chan_laser_2012,
	title = {Laser cooling of an optomechanical crystal resonator to its quantum ground state of motion},
	url = {https://thesis.library.caltech.edu/7098/},
	school = {California Institute of Technology},
	author = {Chan, Jasper},
	month = may,
	year = {2012},
}

@article{kristensen_long-lived_2024,
	title = {Long-lived and {Efficient} {Optomechanical} {Memory} for {Light}},
	volume = {132},
	url = {https://link.aps.org/doi/10.1103/PhysRevLett.132.100802},
	doi = {10.1103/PhysRevLett.132.100802},
	number = {10},
	urldate = {2025-06-10},
	journal = {Physical Review Letters},
	publisher = {American Physical Society},
	author = {Kristensen, Mads Bjerregaard and Kralj, Nenad and Langman, Eric C. and Schliesser, Albert},
	month = mar,
	year = {2024},
	pages = {100802},
}

@article{zeng_phonon_2000,
	title = {Phonon {Heat} {Conduction} in {Thin} {Films}: {Impacts} of {Thermal} {Boundary} {Resistance} and {Internal} {Heat} {Generation}},
	volume = {123},
	issn = {0022-1481},
	shorttitle = {Phonon {Heat} {Conduction} in {Thin} {Films}},
	url = {https://doi.org/10.1115/1.1351169},
	doi = {10.1115/1.1351169},
	number = {2},
	urldate = {2025-05-21},
	journal = {Journal of Heat Transfer},
	author = {Zeng, Taofang and Chen, Gang},
	month = nov,
	year = {2000},
	pages = {340--347},
}

@article{burger_design_2025,
	title = {Design of a release-free piezo-optomechanical quantum transducer},
	volume = {10},
	issn = {2378-0967},
	url = {https://doi.org/10.1063/5.0246075},
	doi = {10.1063/5.0246075},
	number = {1},
	urldate = {2025-04-17},
	journal = {APL Photonics},
	author = {Burger, Paul and Frey, Joey and Kolvik, Johan and Hambraeus, David and Van Laer, Raphaël},
	month = jan,
	year = {2025},
	pages = {010801},
}

@article{komma_thermo-optic_2012,
	title = {Thermo-optic coefficient of silicon at 1550 nm and cryogenic temperatures},
	volume = {101},
	issn = {0003-6951},
	url = {https://doi.org/10.1063/1.4738989},
	doi = {10.1063/1.4738989},
	number = {4},
	urldate = {2024-12-05},
	journal = {Applied Physics Letters},
	author = {Komma, J. and Schwarz, C. and Hofmann, G. and Heinert, D. and Nawrodt, R.},
	month = jul,
	year = {2012},
	pages = {041905},
}

@article{chen_interfacial_2022,
	title = {Interfacial thermal resistance: {Past}, present, and future},
	volume = {94},
	shorttitle = {Interfacial thermal resistance},
	url = {https://link.aps.org/doi/10.1103/RevModPhys.94.025002},
	doi = {10.1103/RevModPhys.94.025002},
	number = {2},
	urldate = {2023-11-26},
	journal = {Reviews of Modern Physics},
	publisher = {American Physical Society},
	author = {Chen, Jie and Xu, Xiangfan and Zhou, Jun and Li, Baowen},
	month = apr,
	year = {2022},
	pages = {025002},
}

@article{meenehan_silicon_2014,
	title = {Silicon optomechanical crystal resonator at millikelvin temperatures},
	volume = {90},
	url = {https://link.aps.org/doi/10.1103/PhysRevA.90.011803},
	doi = {10.1103/PhysRevA.90.011803},
	number = {1},
	urldate = {2023-09-07},
	journal = {Physical Review A},
	publisher = {American Physical Society},
	author = {Meenehan, Seán M. and Cohen, Justin D. and Gröblacher, Simon and Hill, Jeff T. and Safavi-Naeini, Amir H. and Aspelmeyer, Markus and Painter, Oskar},
	month = jul,
	year = {2014},
	pages = {011803},
}

@article{aspelmeyer_cavity_2014,
	title = {Cavity optomechanics},
	volume = {86},
	url = {https://link.aps.org/doi/10.1103/RevModPhys.86.1391},
	doi = {10.1103/RevModPhys.86.1391},
	number = {4},
	urldate = {2023-02-20},
	journal = {Reviews of Modern Physics},
	publisher = {American Physical Society},
	author = {Aspelmeyer, Markus and Kippenberg, Tobias J. and Marquardt, Florian},
	month = dec,
	year = {2014},
	pages = {1391--1452},
}

@article{maccabe_nano-acoustic_2020,
	title = {Nano-acoustic resonator with ultralong phonon lifetime},
	volume = {370},
	issn = {10959203},
	url = {https://www.science.org/doi/10.1126/science.abc7312},
	doi = {10.1126/SCIENCE.ABC7312/SUPPL_FILE/ABC7312_MACCABE_SM.PDF},
	number = {6518},
	urldate = {2022-09-25},
	journal = {Science},
	publisher = {American Association for the Advancement of Science},
	author = {MacCabe, Gregory S. and Ren, Hengjiang and Luo, Jie and Cohen, Justin D. and Zhou, Hengyun and Sipahigil, Alp and Mirhosseini, Mohammad and Painter, Oskar},
	month = nov,
	year = {2020},
	pages = {840--843},
}

@article{borselli_measuring_2006,
	title = {Measuring the role of surface chemistry in silicon microphotonics},
	volume = {88},
	issn = {0003-6951},
	url = {https://aip.scitation.org/doi/abs/10.1063/1.2191475},
	doi = {10.1063/1.2191475},
	number = {13},
	urldate = {2021-10-11},
	journal = {Applied Physics Letters},
	publisher = {American Institute of PhysicsAIP},
	author = {Borselli, Matthew and Johnson, Thomas J. and Painter, Oskar},
	month = mar,
	year = {2006},
	pages = {131114},
}

@article{chan_optimized_2012,
	title = {Optimized optomechanical crystal cavity with acoustic radiation shield},
	volume = {101},
	issn = {0003-6951},
	url = {https://aip.scitation.org/doi/abs/10.1063/1.4747726},
	doi = {10.1063/1.4747726},
	number = {8},
	urldate = {2021-09-06},
	journal = {Applied Physics Letters},
	publisher = {American Institute of PhysicsAIP},
	author = {Chan, Jasper and Safavi-Naeini, Amir H. and Hill, Jeff T. and Meenehan, Seán and Painter, Oskar},
	month = aug,
	year = {2012},
	pages = {081115},
}

\end{document}